\documentclass[10pt,prb,aps,twocolumn,superscriptaddress
]{revtex4-2}

\usepackage{amsmath,amssymb}
\usepackage{newtxtext,newtxmath}
\usepackage{graphicx}
\usepackage{bm}
\usepackage{multirow}
\usepackage{hyperref}

\usepackage{wasysym}
\usepackage{graphicx}
\usepackage{dcolumn}
\usepackage{physics}
\usepackage{mathtools}

\usepackage{algorithm}
\usepackage{algpseudocode}

\usepackage{diagbox}

\usepackage{color}

\definecolor{buff}{rgb}{0.94, 0.86, 0.51}
\definecolor{camel}{rgb}{0.76, 0.6, 0.42}

\newcommand{\Eq}[1]{Eq.~(\ref{#1})}

\usepackage{orcidlink}

\begin{document}
\preprint{APS/123-QED}

\title{Effects of crystal field and momentum-based frustrated exchange interactions
\\ on multiorbital square skyrmion lattice}

\author{Yan~S.~Zha 
\orcidlink{0009-0008-8568-446X}}
\email{yzha@phys.sci.hokudai.ac.jp}
\author{Satoru~Hayami
\orcidlink{0000-0001-9186-6958}}
\email{hayami@phys.sci.hokudai.ac.jp}
\affiliation{Graduate School of Science, Hokkaido University, Sapporo 060-0810, Japan}

\date{\today}
\begin{abstract}
\footnote{This version is content-wise identical to the published version \href{https://journals.aps.org/prb/accepted/10.1103/4sqm-xhw9}{[Yan S. Zha and Satoru Hayami, Phys. Rev. B {\bf 113}, 174415 (2026)]}.}
Motivated by recent theoretical predictions of a square-shaped skyrmion lattice (S-SkL) in
centrosymmetric tetragonal Ce-based magnets
[Yan Zha and Satoru Hayami, \href{https://journals.aps.org/prb/abstract/10.1103/PhysRevB.111.165155}{Phys. Rev. B \textbf{111}, 165155 (2025)}], 
we perform a comprehensive theoretical investigation
into the role of multiorbital effects
and momentum-based frustrated exchange interactions in stabilizing such topologically nontrivial 
magnetic textures.
By employing self-consistent mean-field calculations over a broad range of model parameters, we demonstrate that the cooperative interplay among 
interorbital coupling,
frustrated exchange interactions at higher-harmonic wave vectors, and crystal-field-induced anisotropy is crucial for the stabilization of the S-SkL. 
Furthermore, the competition between the easy-plane intraorbital
anisotropy and the easy-axis interorbital anisotropy leads to a significant enhancement of the S-SkL stability region. 
We also identify 
a rich variety of multi-$Q$ states, including a topologically nontrivial S-SkL state
with a slight breaking of fourfold rotational symmetry (S-SkL$'$), magnetic bubble lattices (MBLs), and double-$Q$ phases with a local/net
scalar chirality.
Our findings elucidate the microscopic mechanism responsible for the emergence of S-SkLs in prototypical Ce-based magnets and provide a route toward realizing skyrmion lattices in a broader class of $f$-electron materials beyond conventional Gd- and Eu-based systems lacking orbital angular momentum.
\end{abstract}

\maketitle
%\tableofcontents

\section{INTRODUCTION} \label{sec:introduction}
Magnetic skyrmions constitute a fascinating platform that 
encompasses three central themes of condensed matter physics---topology, quasiparticles, and symmetry breaking---owing to their nontrivial and unconventional properties~\cite{nagaosa2013topological,zhang2020skyrmion,tokura2020magnetic,gobel2021beyond,hayami2021topological,Zhang_2023review,hayami2024stabilization, kawamura2025frustration,petrovic2025colloquium,koraltan20262026skyrmionicsroadmap}. 
The topological aspect is characterized by a nonzero skyrmion number (topological charge) $n_{\text{sk}}$, defined as 
\begin{align}\label{eq:topoNuminIntro}
    n_{\text{sk}} = \frac{1}{4\pi} \int \bm{n}\cdot \left ( \frac{\partial \bm{n}}{\partial x} \times \frac{\partial \bm{n}}{\partial y} \right ) dx\,dy,
\end{align}
which measures the solid angle subtended by the spin texture within a two-dimensional magnetic unit cell. 
Here, $\bm{n}\equiv\bm{n}(\bm{r})$ denotes the unit-vector field representing the spin direction. 
This topological invariant guarantees the stability of skyrmions against weak perturbations~\cite{zhang2016antiferromagnetic,oike2016interplay,cortes2017thermal,je2020direct}. 
In addition, skyrmions behave as emergent quasiparticles in real space, enabling extensive studies on their manipulation and dynamics~\cite{Jonietzdoi:10.1126/science.1195709,yu2012skyrmion,schulz2012emergent}. 
When conduction electrons traverse the noncoplanar vortex-like texture, their wavefunctions acquire a Berry phase~\cite{berry1984quantal}, giving rise to unconventional transport phenomena such as the topological Hall effect~\cite{Ohgushi_PhysRevB.62.R6065, BrunoPhysRevLett.93.096806,NeubauerPhysRevLett.102.186602,KanazawaPhysRevLett.106.156603,LiPhysRevLett.110.117202}. 
Owing to their nanoscale size, quasiparticle nature, robustness, and transport signatures, skyrmions are considered promising candidates for next-generation information carriers~\cite{romming2013writing,fert2013skyrmions,fert2017magnetic,zhang2020skyrmion}. 

Skyrmions often form long-range ordered phases, commonly referred to as skyrmion lattices (SkLs) or skyrmion crystals (SkXs)~\footnote{In this study, we adopt the former term, SkLs.}. 
The most widely known SkLs appear in noncentrosymmetric magnets where relativistic spin--orbit coupling gives rise to a Dzyaloshinskii--Moriya interaction (DMI)~\cite{DZYALOSHINSKY1958241,MoriyaPhysRev.120.91}. 
In such noncentrosymmetric magnets, SkLs are stabilized by the interplay among DMI, ferromagnetic exchange interaction, and Zeeman coupling to an external magnetic field~\cite{Bogdanov89,BOGDANOV1994255,BinzPhysRevLett.96.207202,roessler2006spontaneous,BinzPhysRevB.74.214408,Muhlbauerdoi:10.1126/science.1166767,SuDoYiPhysRevB.80.054416}.
By contrast, in centrosymmetric systems,
SkLs can also emerge from competing interactions
via the frustration mechanism even without the DMI~\cite{OkuboPhysRevLett.108.017206}. 
Reflecting the distinct stabilization mechanisms, the resulting characteristics of SkLs differ qualitatively between noncentrosymmetric and centrosymmetric systems.  
In 
a frustrated Heisenberg-like effective model under centrosymmetric conditions, the 
\emph{vorticity} and \emph{helicity} degrees of freedom
of the SkL are unlocked owing to the absence of the symmetry-dependent DMI and the bond-dependent anisotropy~\cite{hayamidoi:10.7566/JPSJ.89.103702}
, which renders Bloch-, N\'{e}el-skyrmions and antiskyrmions energetically degenerate~\footnote{In this study, we do not explicitly distinguish between Bloch-, N\'{e}el-skyrmions and antiskyrmions~\cite{ZhaHayamiPhysRevB.111.165155}.}.

In itinerant magnets, competing interactions often arise from the Ruderman--Kittel--Kasuya--Yosida (RKKY) interaction~\cite{RudermanandKittelPhysRev.96.99,Kasuya10.1143/PTP.16.45,YosidaPhysRev.106.893}, expressed as 
\begin{align}
    \mathcal{H}_{\text{RKKY}} = -\mathcal{J}_{\text{K}}^2 \sum_{\bm{q}} \chi_{\bm{q}}^{0}\,\bm{S}_{\bm{q}}\cdot \bm{S}_{-\bm{q}},
\end{align}
where conduction electrons mediate interactions between local moments $\bm{S}_{\bm{q}}$ at wave vector $\pm\bm{q}$ via an exchange coupling $\mathcal{J}_{\text{K}}$. 
At specific nesting wave vectors 
$\bm{Q}_{\eta}$,
the Fermi surface nesting enhances bare susceptibility $\chi_{\bm{q}}^{0}$, which may diverge at low temperature.
Reference~\cite{HayamiPhysRevB.93.184413} pointed out that when $\chi_{\bm{q}}^{0}$ possesses several global maxima at symmetry-related wave vectors 
$\bm{Q}_\eta$,
single-$Q$ helices can evolve into multi-$Q$ states~\cite{MartinPhysRevLett.101.156402,Akagidoi:10.1143/JPSJ.79.083711,KatoPhysRevLett.105.266405,AkagiPhysRevLett.108.096401,BarrosPhysRevB.88.235101,HayamiPhysRevB.90.060402,hayami2021topological,Ozawa_doi:10.7566/JPSJ.85.103703,OzawaPhysRevLett.118.147205}. 
This mechanism has been experimentally confirmed in centrosymmetric itinerant magnets such as Gd$_2$PdSi$_{3}$ with a triangular lattice~\cite{DongPhysRevLett.133.016401} and GdRu$_2$Si$_2$ with a tetragonal lattice~\cite{Dongdoi:10.1126/science.adj7710}.

In insulators, even short-range exchange interactions take on the role of the RKKY interaction by considering further neighbor contributions. 
Within the Luttinger-Tisza framework~\cite{LITVIN1974205}, by
maximizing the Fourier-transformed exchange interaction $\mathcal{J}_{\bm{q}}$ (or equivalently minimizing $-\mathcal{J}_{\bm{q}}$), multiple degenerate global maxima $\{ \bm{Q}_\eta\}$ can be realized, and the existence of symmetry-allowed maxima provides 
a
favorable condition for multi-$Q$ states~\cite{HayamiPhysRevB.93.184413}. 
We emphasize, however, that this criterion identifies only candidate ordering wave vectors at the 
quadratic instability level. 
Stabilizing noncollinear/noncoplanar multi-$Q$ states (e.g., SkL) generally requires additional ingredients that generate effective higher-order terms in the free energy~\cite{OkuboPhysRevLett.108.017206,UtesovPhysRevB.103.064414, UtesovPhysRevB.105.054435,HayamiPhysRevB.94.174420,amoroso2020spontaneous,HayamiMotomePhysRevB.103.054422,YambePhysRevB.106.174437,ZhangLinPhysRevLett.133.196702,YambePhysRevB.110.014428,ShiratoJPSJ.94.063601}.

In this study, we focus on the square-shaped
SkL (S-SkL), which was experimentally observed in the centrosymmetric magnet GdRu$_2$Si$_2$~\cite{khanh2020nanometric,yasui2020imaging,khanh2022zoology,Wood_PhysRevB.107.L180402}.
The S-SkL is
constructed from a superposition of two perpendicular wave vectors $\bm{Q}_1$ and $\bm{Q}_2$ along the $\langle 110\rangle$ or $\langle100\rangle$ directions, possessing fourfold rotational symmetry on a square lattice without geometrical frustration. 
The S-SkL, consisting of double-$Q$ spin density waves, is distinct from the triangular SkL (T-SkL), consisting of triple-$Q$ spin density waves at $\bm{Q}_1$, $\bm{Q}_2$, and $\bm{Q}_3$, from the energetic viewpoint.
The latter T-SkL is naturally stabilized by the quartic coupling $(\bm{S}_{\bm{0}} \cdot \bm{S}_{\bm{Q}_1})(\bm{S}_{\bm{Q}_2} \cdot \bm{S}_{\bm{Q}_3})$ among three symmetry-equivalent helices and the uniform magnetization owing to the relation of $\bm{Q}_1+\bm{Q}_2+\bm{Q}_3=\bm{0}$. 
The stabilization of the S-SkL, however,
is difficult
within the frustrated Heisenberg-like effective model under the magnetic field alone~\cite{OkuboPhysRevLett.108.017206,SZLinPhysRevB.93.064430,HayamiPhysRevB.93.184413}, since the double-$Q$ superposition on a square lattice lacks the 
momentum-closure
condition (e.g., $\bm{Q}_1+\bm{Q}_2 \neq \bm{0}$), indicating the small quartic coupling. 
To achieve the stabilization of the S-SkL,
additional interactions are introduced, such as
a higher-order, perturbatively generated biquadratic interaction in itinerant magnets~\cite{Hayamidoi:10.7566/JPSJ.91.023705, HayamiPhysRevB.103.024439}, compass-type anisotropy~\cite{UtesovPhysRevB.103.064414, ZTWangPhysRevB.103.104408}, higher-harmonic couplings~\cite{ZTWangPhysRevB.103.104408,Hayamidoi:10.7566/JPSJ.91.023705,HayamiPhysRevB.105.174437,HAYAMI2023170547,OkigamiPhysRevB.110.L220405,Hayamimagnetism5020012}.
These theoretical studies have
elucidated or predicted the microscopic mechanisms of S-SkLs, providing substantial theoretical support for experimental studies on S-SkLs.
Especially, the higher-harmonic couplings at $\bm{Q}'_2 \equiv \bm{Q}_1+\bm{Q}_2$
can naturally lead to the instability toward S-SkLs
when their magnitude becomes comparable to that of the interactions at $\bm{Q}_1$ and $\bm{Q}_2$, since a situation satisfying the 
momentum-closure condition $\bm{Q}_1+\bm{Q}_2-(\bm{Q}'_2)=\bm{0}$ can be realized.

Furthermore, it is noteworthy that most known centrosymmetric S-SkL materials involve $4f$ lanthanoid elements with a quenched orbital angular momentum, such as Gd and Eu ions~\cite{kurumaji2019skyrmion,hirschberger2019skyrmion,khanh2020nanometric,takagi2022square,yoshimochi2024multistep}, while 
$f$-electron compounds with finite orbital angular momentum have not been extensively explored. 
Moreover, most theoretical studies have focused only on spin degrees of freedom, neglecting orbital contributions. 
This motivates 
us to investigate multiorbital effects in S-SkLs. 

To this end, we investigate the instability toward the S-SkL state arising from the synergy between crystal-field effects, containing magnetic anisotropy and interorbital coupling, and momentum-based frustrated exchange interactions at higher-harmonic wave vectors in multiorbital systems.
Our previous study Ref.~\cite{ZhaHayamiPhysRevB.111.165155} established the minimal ingredients required for stabilizing the S-SkL in $f$-electron systems.
In the present work, we go beyond that baseline in four key respects:
(i)~a microscopic analysis of the relevant matrix elements of the total angular-momentum operator $\hat{\bm{J}}$ within and between the two Kramers doublets, clarifying how the crystal-field orbital wave functions' superposition coefficient $\alpha$ controls the competition between intraorbital anisotropy and interorbital coupling (Sec.~\ref{subsec:alpha});
(ii)~an explicit energy decomposition and diagnostics to disentangle the contributions from the two orbital subspaces and their coupling channel [Eqs.~(\ref{eq:Htot_decomposed}) and~(\ref{eq:Utot_decomposed})], identifying the energetic origin and parameter window of the S-SkL (Sec.~\ref{subsec:alpha});
(iii)~a controlled test of interorbital coupling strength via a
weighting coefficient $\gamma$ on the off-diagonal blocks (Sec.~\ref{subsec:weighting});
and (iv)~a controlled comparison of higher-harmonic regimes via a
dimensionless ratio $\xi$, demonstrating the role of momentum-space
frustration in stabilizing the S-SkL (Sec.~\ref{subsec:higherharmonics}).
These results, obtained by systematically varying the three
microscopic parameters ($\alpha,\,\gamma,\,\xi$), yield a set of mechanism-driven phase diagrams and provide design principles for $4f$-electron tetragonal magnets (e.g., Ce-based compounds) with unquenched orbital angular momentum.

The rest of the paper is organized as follows.
In Sec.~\ref{sec:model}, we introduce the crystal-field orbital states (two $\Gamma_{t7}$ Kramers doublets) and an effective localized model incorporating the exchange interaction, Zeeman coupling, and the signed crystal-field splitting $\Delta$.
In Sec.~\ref{sec:results}, we present several low-temperature phase diagrams and discuss how magnetic anisotropy, interorbital coupling, and higher-harmonic wave vectors govern the stability of the S-SkL and accompanying states, including magnetic bubble lattices (MBLs).
In Sec.~\ref{sec:discussion}, we provide a summary of the main findings and offer brief remarks on potential experimental probes.
Finally, in Sec.~\ref{sec:summary}, we conclude and discuss material implications.
To maintain a self-contained presentation while avoiding redundancy with Ref.~\cite{ZhaHayamiPhysRevB.111.165155}, we relegate technical details to the Appendices:
Appendices~\ref{app:CEF} and~\ref{app:Jxyz} present the crystal-field Hamiltonian and the explicit forms of the projected total-angular-momentum operators, respectively.
Appendix~\ref{app:computational methodology} describes the self-consistent mean-field procedure.
Appendix~\ref{app:physical quantities} summarizes the physical quantities, and Appendix~\ref{app:phaseclassification} details the numerical accuracy and the criteria used for phase classification using the physical quantities.
Appendix~\ref{app:states} provides snapshots of the eight representative states documented in Ref.~\cite{ZhaHayamiPhysRevB.111.165155}.
Appendix~\ref{app:tensor_exchange} briefly discusses the possibility of multipolar (tensorial) exchange beyond the bilinear form.

\section{MODEL} \label{sec:model}
We consider a
localized $4f$-electron system on a two-dimensional square lattice, with one magnetic Ce$^{3+}$ ion (having an outermost $f^1$ electron configuration) at each lattice site. Owing to the typically strong relativistic spin--orbit coupling
compared with the crystal-field effects in lanthanoid compounds, the fourteenfold-degenerate $4f$ manifold splits into an eightfold-degenerate $\ket{7/2, J^z}$ multiplet and a sixfold-degenerate $\ket{5/2, J^z}$ multiplet~\cite{yosida1996theory}.
The latter, conventionally denoted as $^2F_{5/2}$, represents the energetically lower states; in the following, we consider their further splitting by introducing the tetragonal crystalline electric field.

The crystal field can be expanded in terms of spherical harmonics, corresponding to the multipole expansion in classical electromagnetism~\cite{Griffiths_2023}.
In the modern multipole description, the spin--orbital--charge degrees of freedom can be classified into four types of multipoles with rank $l$~\cite{hayami2024unified}: electric multipole $(\mathcal{P}, \mathcal{T}) = 
[(-1)^l, +1]$, magnetic multipole $(\mathcal{P}, \mathcal{T}) = 
[(-1)^{l+1}, -1]$, electric toroidal multipole $(\mathcal{P}, \mathcal{T}) = 
[(-1)^{l+1}, +1]$, and magnetic toroidal multipole $(\mathcal{P}, \mathcal{T}) = 
[(-1)^l, -1]$, which differ in their $\mathcal{P}$ and $\mathcal{T}$ symmetries. Here, $\pm 1$ denotes whether the corresponding symmetry is
present ($+1$) or 
absent ($-1$), with $\mathcal{P}$ representing spatial inversion (parity) symmetry and $\mathcal{T}$ representing time-reversal symmetry.
Due to the totally symmetric representation $A_{1g}$ of the $D_{\rm 4h}$ point group---a typical symmetry of tetragonal Ce-based compounds---only the rank-2 electric quadrupole and rank-4 electric hexadecapole are allowed among the electric multipoles~\cite{Kusunosedoi:10.1143/JPSJ.77.064710}.
These electric multipoles contribute to the crystal-field Hamiltonian $\hat{\mathcal{H}}_{\text{CEF}}$~\cite{stevens1952matrix,hutchings1964point},
and a detailed form of $\hat{\mathcal{H}}_{\text{CEF}}$ has been shown in Appendix~\ref{app:CEF}.

The $D_{\rm 4h}$ symmetry of tetragonal Ce-based compounds further splits the Ce$^{3+}$
ground state $\ket{5/2,J^z}$ into three Kramers doublets: two $\Gamma_{t7}$ doublets  
$\ket{\Gamma_{t7\pm}^{(1)}} 
        = \alpha \ket{\tfrac{5}{2},\pm \tfrac{5}{2}} 
          - \beta \ket{\tfrac{5}{2},\mp \tfrac{3}{2}}$ and  
$\ket{\Gamma_{t7\pm}^{(2)}} 
        = \beta \ket{\tfrac{5}{2},\pm \tfrac{5}{2}} 
          + \alpha \ket{\tfrac{5}{2},\mp \tfrac{3}{2}}$,  
and one $\Gamma_{t6}$ doublet  
$ \ket{\Gamma_{t6\pm}} = \ket{\tfrac{5}{2},\pm \tfrac{1}{2}}$~\cite{sato2013appendixB,SundermannPhysRevB.99.235143,ZhaHayamiPhysRevB.111.165155},  
where $\alpha$ and $\beta$ originate from crystal-field parameters (Appendix~\ref{app:CEF}) and $\alpha^2 + \beta^2 = 1$ ensures the normalization of the crystal-field orbital wave functions.
We consider two low-lying $\Gamma_{t7}$ Kramers doublets, $\ket{\Gamma_{t7\pm}^{(1)}}$ and $\ket{\Gamma_{t7\pm}^{(2)}}$.
The relative level ordering of the two $\Gamma_{t7}$ Kramers doublets is controlled by the signed on-site level offset $\Delta$ in $\hat{\mathcal{H}}_{\Delta}$, introduced below~\footnote{Throughout this paper, $\Delta$ is treated as a signed level-offset parameter: $|\Delta|$ sets the magnitude of the crystal-field splitting, while $\mathrm{sign}(\Delta)$ specifies the bare ordering of the two doublets.}.

In addition to the local term, the many-body exchange interaction is introduced in a simplified bilinear form: $\mathcal{J} \bm{J}_i \cdot \bm{J}_j$, where $\mathcal{J}$ denotes an
exchange interaction
\footnote{We suppose that the anisotropic effect arises from the local crystalline electric field, which also affects the anisotropy of the exchange interaction.}.
The exchange Hamiltonian is given by
\begin{align}\label{eq:H_ex}
    \hat{\mathcal{H}}_{\text{ex}} = -\sum_{\langle i,j \rangle} \mathcal{J}_{ij} \hat{\bm{J}}_i \cdot \hat{\bm{J}}_j,
\end{align}
where $\hat{\bm{J}}_{i (j)}$ represents the localized total angular-momentum operator at site $i (j)$.
Projecting the low-energy $J=5/2$ manifold further onto the two $\Gamma_{t7}$ Kramers doublets yields an effective $4$-dimensional basis $\left\{ \ket{\Gamma_{t7 +}^{(1)}} = \ket{1}, \ket{\Gamma_{t7 -}^{(1)}}= \ket{2},\ket{\Gamma_{t7 +}^{(2)}}= \ket{3}, \ket{\Gamma_{t7 -}^{(2)}} = \ket{4}  \right\}$ ($\ket{1}$--$\ket{4}$ stands for the short notation), in which the explicit $4\times4$ matrices of $\hat{J}^{x,y,z}$ are given in Appendix~\ref{app:Jxyz}.

By employing the mean-field approximation for the interaction term~\cite{weiss1907hypothese}, the total effective Hamiltonian is written as
\begin{eqnarray} \label{eq:TotMFhamiltonian}
    \hat{\mathcal{H}}_{\text{tot}} &=& \hat{\mathcal{H}}^{\text{MF}}_{\text{ex}} + \hat{\mathcal{H}}_{\text{Z}} + 
    \hat{\mathcal{H}}_{\Delta}, \\
    \hat{\mathcal{H}}^{\text{MF}}_{\text{ex}} 
    &=& - \sum_{\langle i,j\rangle} \mathcal{J}_{ij} \big( 
        \hat{\bm{J}}_i \cdot \langle \hat{\bm{J}}_j \rangle + \hat{\bm{J}}_j \cdot \langle \hat{\bm{J}}_i \rangle -
       \langle \hat{\bm{J}}_i \rangle \cdot \langle \hat{\bm{J}}_j \rangle \big), \label{eq:H_MFex} \\
    \hat{\mathcal{H}}_{\text{Z}} &=& -h \sum_{i} \hat{J}_{i}^{z},  \\
    \hat{\mathcal{H}}_{\Delta} &=& 
      \Delta \sum_i 
       \hat{P}_{34},
       \label{eq:H_Delta} 
\end{eqnarray}
where $\hat{\mathcal{H}}^{\text{MF}}_{\text{ex}}$ describes the mean-field-approximated exchange interactions, $\hat{\mathcal{H}}_{\text{Z}}$ is the Zeeman coupling to the external magnetic field $h$ along the $z$ axis, 
and $\hat{\mathcal{H}}_{\Delta}$ represents the crystal-field splitting between the two $\Gamma_{t7}$ Kramers doublets with magnitude $\abs{\Delta}$ and sign specifying the 
bare
level ordering.
Here, $\hat{P}_{34}\equiv \ket{3}_i\bra{3}_i+\ket{4}_i\bra{4}_i$ is the projector onto the $(3,4)$ Kramers doublet at site $i$.
Similarly, $\hat{P}_{12}\equiv \ket{1}_i\bra{1}_i+\ket{2}_i\bra{2}_i$ projects onto the $(1,2)$ doublet.
Using $\hat{P}_{12}+\hat{P}_{34}=\mathbb{I}$, one has $-|\Delta|\hat{P}_{34}=-|\Delta|\mathbb{I}+|\Delta|\hat{P}_{12}$, i.e., $\Delta<0$ is equivalent up to an additive constant (and relabeling) within the two-doublet subspace.
The constant shift does not affect any physical observable.

Next, we discuss the nature of the exchange interaction in \Eq{eq:H_ex} from the momentum-space viewpoint.
By performing the Fourier transformation on the exchange Hamiltonian in Eq.~(\ref{eq:H_ex}), $- \sum_{\bm{q}} \hat{\bm{J}}_{\bm{q}} \cdot \hat{\bm{J}}_{-\bm{q}} \,\mathcal{J}(\bm{q})$ with a Fourier-transformed
exchange interaction $\mathcal{J}(\bm{q}) = 2\mathcal{J}_1 \left( \cos{q_x} + \cos{q_y} \right) 
+ 4\mathcal{J}_2\, \cos{q_x} \cos{q_y} 
+ 2\mathcal{J}_3 \left( \cos{2q_x} + \cos{2q_y} \right)$ for the square lattice with the interactions $\mathcal{J}_1$--$\mathcal{J}_2$--$\mathcal{J}_3$~\cite{ZhaHayamiPhysRevB.111.165155}. 
We set the lattice constant $a = 1$ in this study and choose short-period ordering wave vectors $\bm{Q}_{1} = \left(2\pi/6,2\pi/6 \right)$ and $\bm{Q}_{2} = \left(-2\pi/6,2\pi/6 \right)$ (and their negatives),
which lead to $6\times 6$-site ordered states at most, allowing 
us to reduce computational costs.
For isotropic interactions in the single-orbital model, the ground state is a helical state with either $\bm{Q}_1$ or $\bm{Q}_2$~\cite{AYoshimoridoi:10.1143/JPSJ.14.807}.
The introduction of magnetic anisotropy in cooperation with the Zeeman coupling can lead to the multi-$Q$ instability, as demonstrated in the spin-only model~\cite{ZTWangPhysRevB.103.104408}. 

Although infinitely many possible combinations of $\{\mathcal{J}_2, \mathcal{J}_3 \}$ can 
yield the ordering wave vector at $\bm{Q}_1$, previous studies have shown that higher-harmonic wave vectors are crucial for stabilizing the S-SkL, MBLs, and other trivial 
2$Q$ states on the square lattice in centrosymmetric frustrated systems without DMI~\cite{ZTWangPhysRevB.103.104408,Hayamidoi:10.7566/JPSJ.91.023705,HayamiPhysRevB.105.174437,HAYAMI2023170547,ZhaHayamiPhysRevB.111.165155,Hayamimagnetism5020012}. 
Therefore, it is important to choose $\{\mathcal{J}_2, \mathcal{J}_3 \}$ such that the contribution from higher harmonics can be systematically tuned. 
In the present study, we follow the approach of Refs.~\cite{ZTWangPhysRevB.103.104408,HayamiPhysRevB.105.174437}. 
In multi-$Q$ states, magnetic textures are modulated by two perpendicular fundamental ordering wave vectors, while higher harmonics arise from integer linear combinations of these fundamentals, such as $\bm{Q}_1 - \bm{Q}_2 \equiv \bm{Q}'_1$ and $\bm{Q}_1 + \bm{Q}_2 \equiv \bm{Q}'_2$. 
To quantify their contribution, we define a dimensionless ratio~\cite{HayamiPhysRevB.105.174437}
\begin{align}\label{eq:xi}
    \xi = \frac{\mathcal{J}_{\bm{Q}'_{\eta}}}{\mathcal{J}_{\bm{Q}_\eta}},\quad \eta \in \{1,2\},
\end{align}
to quantify higher harmonics' relative weight compared with the dominant modes.
Here, we
adopt $\xi = 0.875$ ($\mathcal{J}_2=-0.5, \mathcal{J}_3=-0.25$) and $0.268$ ($\mathcal{J}_2=-0.1, \mathcal{J}_3=-0.45$) corresponding to strong and weak higher-harmonic contributions, respectively.

\section{RESULTS} \label{sec:results}
In this section, we first examine the $\alpha$ dependence over a wider range of the crystal-field splitting $\Delta$,
thereby extending the analysis presented in Ref.~\cite{ZhaHayamiPhysRevB.111.165155} (see Sec.~\ref{subsec:alpha}). 
Next, a weighting coefficient $\gamma$ is introduced at fixed $\alpha$ and $\Delta$ to further investigate the effect of interorbital coupling 
(Sec.~\ref{subsec:weighting}). 
Finally, we demonstrate that higher-harmonic wave vectors play a crucial role in stabilizing the S-SkL and other double-$Q$ magnetic states in this system (Sec.~\ref{subsec:higherharmonics}). 
A detailed description of the self-consistent mean-field procedure is provided in Appendix~\ref{app:computational methodology}.
We also provide several physical quantities consistent with Ref.~\cite{ZhaHayamiPhysRevB.111.165155} used for the phase classification in Appendix~\ref{app:physical quantities}.
The numerical accuracy and criteria used to classify the obtained magnetic-moment configurations are described in Appendix~\ref{app:phaseclassification}.

\subsection{Effects of magnetic anisotropy}
\label{subsec:alpha}

\begin{figure}[htbp]
    \centering
    \includegraphics[width=8.5cm]{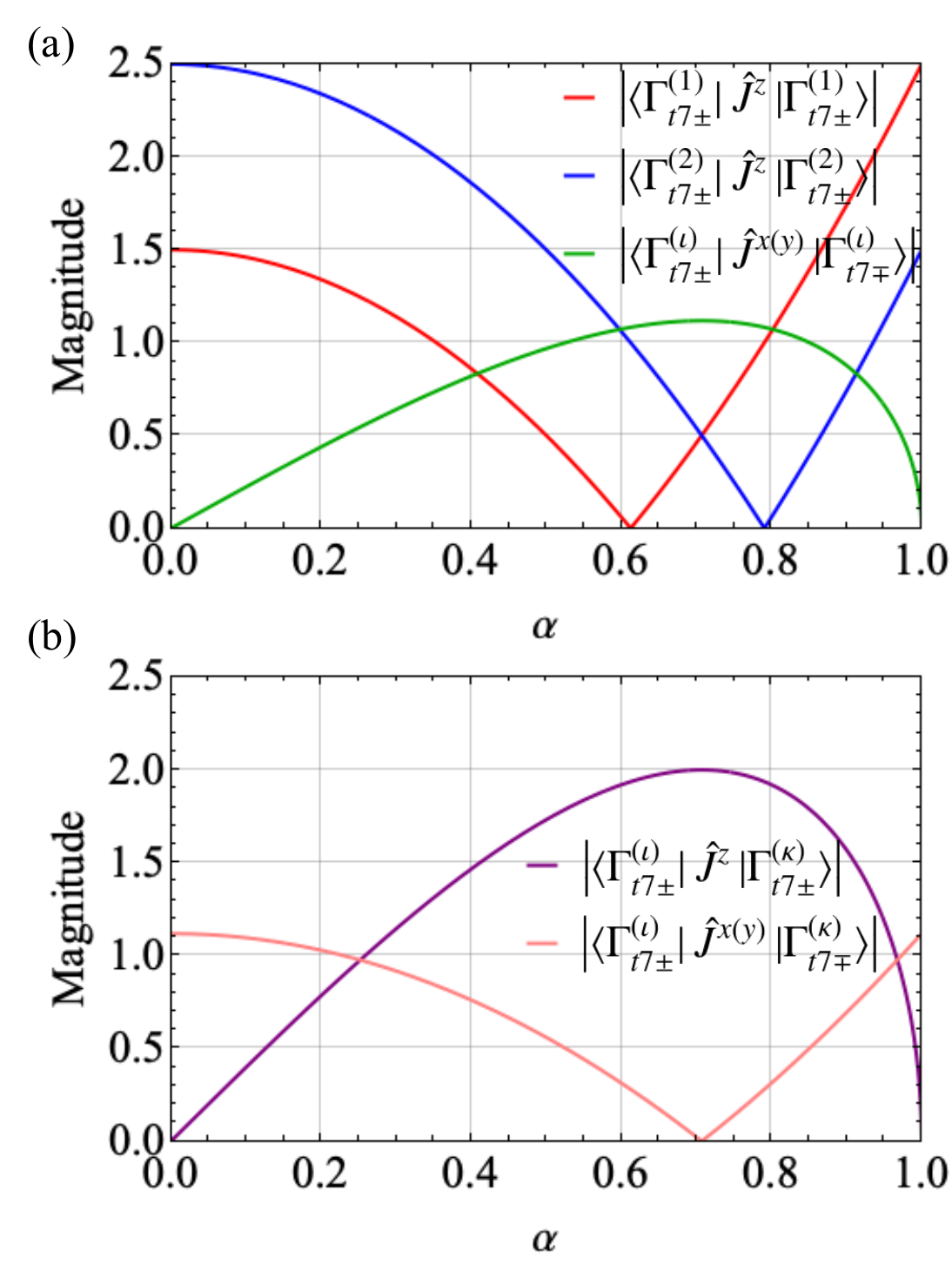}
    \caption{(a) Dependence on the superposition coefficient $\alpha$ ($0 \le \alpha \le 1$) of $\abs{\bra{\Gamma_{t7 \pm}^{(1)}}\hat{J}^z\ket{\Gamma_{t7 \pm}^{(1)}}}$, $\abs{\bra{\Gamma_{t7 \pm}^{(2)}}\hat{J}^z\ket{\Gamma_{t7 \pm}^{(2)}}}$, and $\abs{\bra{\Gamma_{t7\pm}^{(\iota)}} \hat{J}^{x(y)}\ket{\Gamma_{t7\mp}^{(\iota)}}}$ with $\iota \in \{1,2\}$. The first two correspond to the diagonal elements of $J^z$ for the two Kramers doublets, while the last corresponds to the off-diagonal elements of $\hat{J}^x$ (and $\hat{J}^y$). (b) $\alpha$ dependence of $\abs{\bra{\Gamma^{(\iota)}_{t7\pm}}\hat{J}^z \ket{\Gamma^{(\kappa)}_{t7\pm}}}$ and $\abs{\bra{\Gamma^{(\iota)}_{t7\pm}}\hat{J}^{x(y)} \ket{\Gamma^{(\kappa)}_{t7\mp}}}$ with $\iota,\kappa\in\{1,2\},\ \iota\neq\kappa$. These off-diagonal matrix elements of $\hat{J}^z$ and $\hat{J}^{x(y)}$ encode the interorbital coupling between the two $\Gamma_{t7}$ Kramers doublets.}
    \label{fig_alpha}
\end{figure}

The degree of magnetic anisotropy induced by the crystal-field orbital wave functions can be evaluated by comparing the magnitudes of the relevant matrix elements of the projected operators $\hat{J}^{x,y,z}$.
Specifically, these are given by
\begin{align}\label{eq:a1}
    &\abs{\bra{\Gamma_{t7 \pm}^{(1)}}\hat{J}^z\ket{\Gamma_{t7 \pm}^{(1)}}} = \abs{\tfrac{1}{2}(5 \alpha^2 - 3 \beta^2)}, \nonumber \\
    &\abs{\bra{\Gamma_{t7 \pm}^{(2)}}\hat{J}^z\ket{\Gamma_{t7 \pm}^{(2)}}} = \abs{\tfrac{1}{2}(5 \beta^2 - 3 \alpha^2)},  \\
    &\abs{\bra{\Gamma_{t7\pm}^{(\iota)}} \hat{J}^{x(y)} \ket{\Gamma_{t7\mp}^{(\iota)}}} = \sqrt{5}\alpha \beta, \qquad  \iota \in \{1,2\}. \nonumber
\end{align}
The first two correspond to the diagonal elements of $\hat{J}^z$ in the two $2\times 2$ Kramers subspaces, while the third corresponds to the magnitude of the off-diagonal elements of $\hat{J}^x$ (and equivalently $\hat{J}^y$) in each subspace.
Similarly, the magnitudes of the matrix elements in the off-diagonal blocks, which encode the interorbital coupling, are expressed as
\begin{align}\label{eq:a2}
    &\abs{\bra{\Gamma^{(\iota)}_{t7\pm}}\hat{J}^z \ket{\Gamma^{(\kappa)}_{t7\pm}}} = 4\alpha \beta, \nonumber\\
    &\abs{\bra{\Gamma^{(\iota)}_{t7\pm}}\hat{J}^{x(y)} \ket{\Gamma^{(\kappa)}_{t7\mp}}} = \abs{ \frac{1}{2} \sqrt{5} (\alpha -\beta ) (\alpha +\beta )},  \\
    &\qquad \qquad \qquad \qquad \qquad \iota,\kappa\in\{1,2\},\ \iota\neq\kappa. \nonumber 
\end{align}
Figures~\ref{fig_alpha}(a)~and~\ref{fig_alpha}(b) present the $\alpha$ dependence 
obtained from Eqs.~(\ref{eq:a1})~and~(\ref{eq:a2}), respectively.

In this study, we take $\alpha = 0.3, 0.38, 0.408, 0.6124,$ and $0.65$ to investigate the stability tendency of the S-SkL against the magnetic anisotropy.  
Summary of the anisotropy characteristics of the two $\Gamma_{t7}$ Kramers doublets as a function of the superposition coefficient $\alpha$ is shown in Table~\ref{table:alpha_anisotropy}.  
The classification is based on the magnitudes of the diagonal and off-diagonal matrix elements in each $2 \times 2$ block of $\hat{\bm{J}}$ shown in Fig.~\ref{fig_alpha}, corresponding to Eqs.~(\ref{eq:a1})~and~(\ref{eq:a2}).  
``Easy-axis'' and ``easy-plane'' denote predominant anisotropy along the $z$ axis and within the $xy$ plane, respectively, while ``isotropic'' indicates nearly equal components in all directions.
For $\alpha = 0.3$ and $0.38$, both $\Gamma_{t7}$ doublets exhibit easy-axis anisotropy along the $z$ axis, with the $\ket{\Gamma_{t7 \pm}^{(2)}}$ doublet showing a particularly strong anisotropy.  
At $\alpha = 0.408$, $\ket{\Gamma_{t7 \pm}^{(1)}}$ becomes isotropic, while $\ket{\Gamma_{t7 \pm}^{(2)}}$ retains a strong $z$-axis easy-axis anisotropy.  
For $\alpha = 0.6124$ and $0.65$, both doublets exhibit easy-plane anisotropy in the $xy$-plane; notably, at $\alpha = 0.6124$, the $\hat{J}^z$ diagonal elements for $\ket{\Gamma_{t7 \pm}^{(1)}}$ vanish.  
It should be noted that the magnetic anisotropy of the system is determined not only by this intraorbital-space anisotropy but also by the interorbital coupling, which corresponds to the $2 \times 2$ off-diagonal block elements in the $4\times4$ matrix representation of $\hat{\bm{J}}$ as shown in Fig.~\ref{fig_alpha}(b) and \Eq{eq:a2}.
In the present setup, the 
interorbital coupling is characterized by easy-axis anisotropy for all chosen $\alpha$.

\begin{figure*}[t!]
    \begin{center}
    \includegraphics[width=1.0 \hsize]{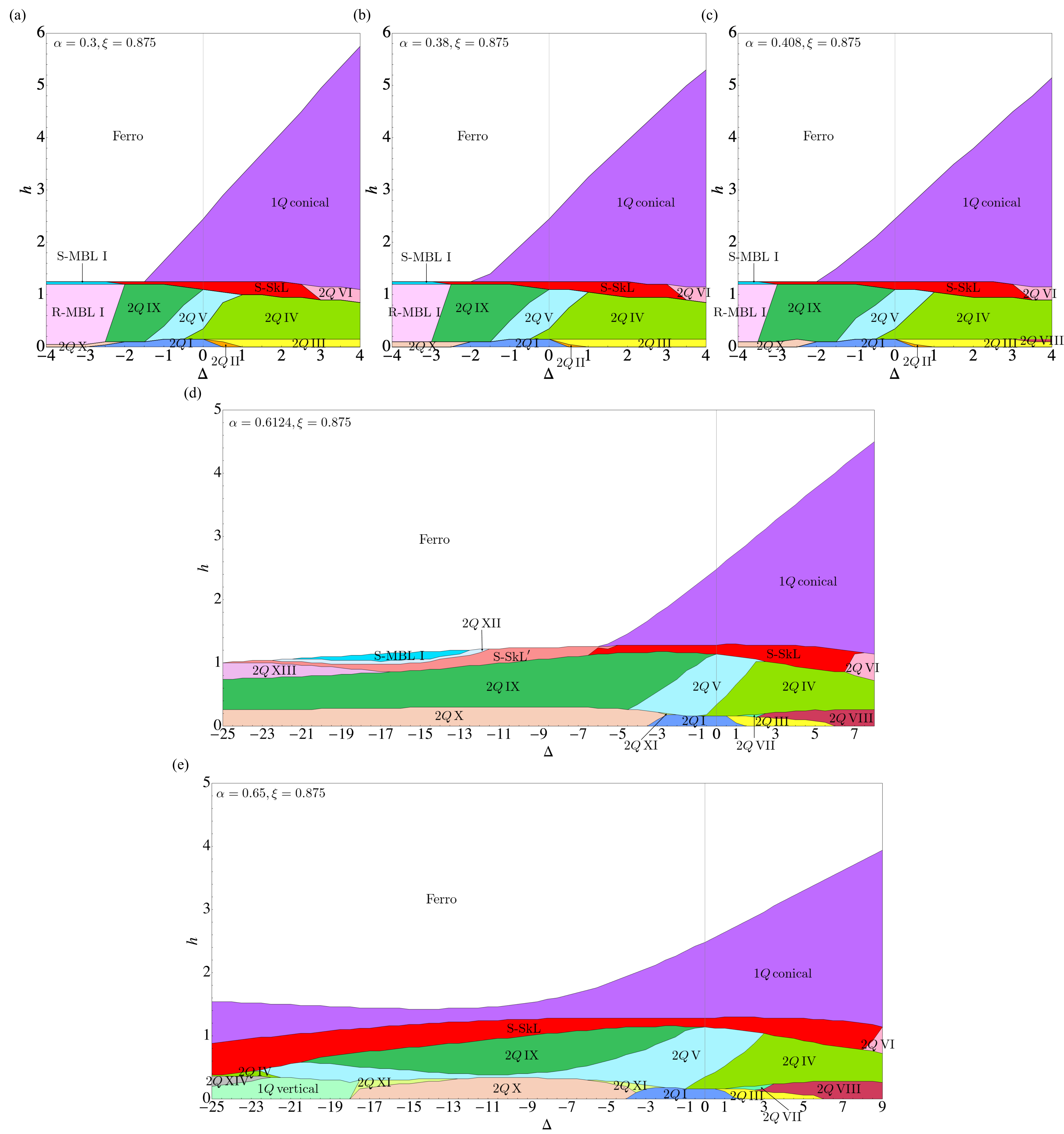}
    \caption{(a)--(e) $\Delta$--$h$ phase diagrams for $\alpha = 0.3$, $0.38, 0.408, 0.6124$, and $0.65$
    at low temperature ($T = 0.05$) in the square-lattice system with the relatively strong higher-harmonic wave-vector
    contribution ($\xi = 0.875$). 
    The magnetic field $h$ and the crystal-field splitting $\Delta$ (between the two $\Gamma_{t7}$ Kramers doublets) are shown on the vertical and horizontal axes, respectively. 
    The negative-$\Delta$ region corresponds to a reversal of bare level ordering.
    The crossover boundaries of the two Kramers doublets in the interacting system are generally shifted away from $\Delta=0$; see Fig.~\ref{fig_delta_energy}.
    Each colored region represents a distinct low-temperature magnetic configuration, including the 1$Q$~CS state, 1$Q$~VS state, various 2$Q$ states (labeled I--XIV), the S-MBL~I and R-MBL~I states, the S-SkL and S-SkL$'$ states, and the fully polarized ferromagnetic (Ferro) state.
    }
    \label{fig_phasedia_alpha}
    \end{center}
\end{figure*}

\begin{table}[h!]
\centering
\caption{Summary of the anisotropy characteristics of the two $\Gamma_{t7}$ Kramers doublets as a function of the superposition coefficient $\alpha$.
}
\label{table:alpha_anisotropy}
\begin{tabular}{c|ccc}
\hline\hline
\diagbox{$\alpha$}{$\Gamma_{t7}$} &  $\ket{\Gamma_{t7 \pm}^{(1)}}$ & $\ket{\Gamma_{t7 \pm}^{(2)}}$ & interorbital coupling\\
\hline
$0.3$   & \text{easy-axis} & \text{easy-axis} & \text{easy-axis} \\
$0.38$  & \text{easy-axis} & \text{easy-axis} & \text{easy-axis} \\
\hline
$0.408$ & \text{isotropic} & \text{easy-axis} & \text{easy-axis} \\
\hline
$0.6124$ & \text{easy-plane} & \text{easy-plane} & \text{easy-axis} \\
$0.65$  & \text{easy-plane} & \text{easy-plane} & \text{easy-axis} \\ \hline 
\hline
\end{tabular}
\end{table}
Figures~\ref{fig_phasedia_alpha}(a)--\ref{fig_phasedia_alpha}(e) show the $\Delta$--$h$ phase diagrams for $\alpha = 0.3, 0.38, 0.408, 0.6124$, and $0.65$, respectively, at a low temperature of $T = 0.05$ and 
at the higher-harmonic wave-vector 
contribution of $\xi = 0.875$, obtained from iterative mean-field calculations. 
We extend the results of Ref.~\cite{ZhaHayamiPhysRevB.111.165155} at $\alpha=0.38$ by enlarging the $\Delta$ range from $-4$ to $4$, and we also depict the phase diagrams for $\alpha=0.30$ and $0.408$ over the same $\Delta$ window, as shown in Figs.~\ref{fig_phasedia_alpha}(a)--\ref{fig_phasedia_alpha}(c).
Furthermore, we show an even wider range of $\Delta$, $-25 \le \Delta \le 8$ and $-25 \le \Delta \le 9$, for $\alpha = 0.6124$ in Fig.~\ref{fig_phasedia_alpha}(d) and $\alpha = 0.65$ in Fig.~\ref{fig_phasedia_alpha}(e), respectively.
The magnetic-field interval is set to $0.05$ for $\alpha = 0.3$, $0.38$, and $0.408$, and to $0.02$ for $\alpha = 0.6124$ and $0.65$. 
The interval of the crystal-field splitting $\Delta$ is fixed at $0.5$.
The phase boundaries are obtained by smoothly interpolating the discrete data points.

\begin{figure}[htbp]
    \begin{center}
    \includegraphics[width=8.0cm]{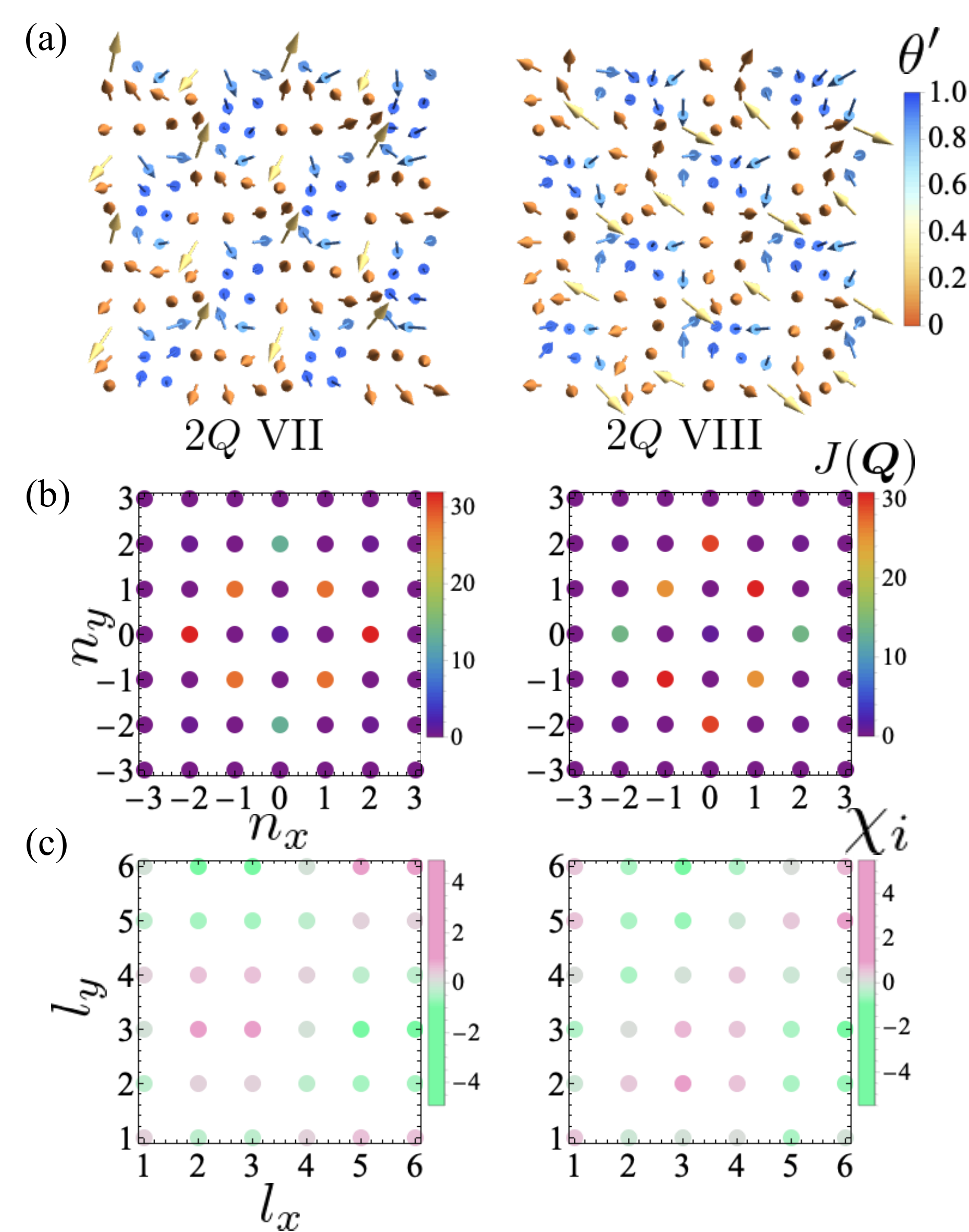}
    \caption{
    Real-space and momentum-space characteristics of the magnetic phases 2$Q$~VII and 2$Q$~VIII stabilized either only in $\Delta>0$ or in both $\Delta>0$ and $\Delta<0$ regions, depending on parameters.
    Panels (a)--(c) show (a) the three-dimensional magnetic-moment configurations in a $6 \times 6$ unit cell, where the magnetic moments are drawn at each site and normalized in length for clarity; 
    the color scale represents the normalized polar angle $\theta' = \theta / \pi$, varying continuously from 0 (north pole) to 1 (south pole); 
    (b) the structure-factor distributions $J(\bm{q})$ in momentum space, where $n_x$ and $n_y$ denote multiples of $2\pi / 6$ with $-3 \le n_x, n_y \le 3$ in the first Brillouin zone; 
    and (c) the local scalar chirality $\chi_i$ within the $6 \times 6$ unit cell.
    }
    \label{fig_RealspaceSfChirality_alpha_plus_78}
    \end{center}
\end{figure}

\begin{figure*}[t!]
    \centering
    \includegraphics[width=1\hsize]{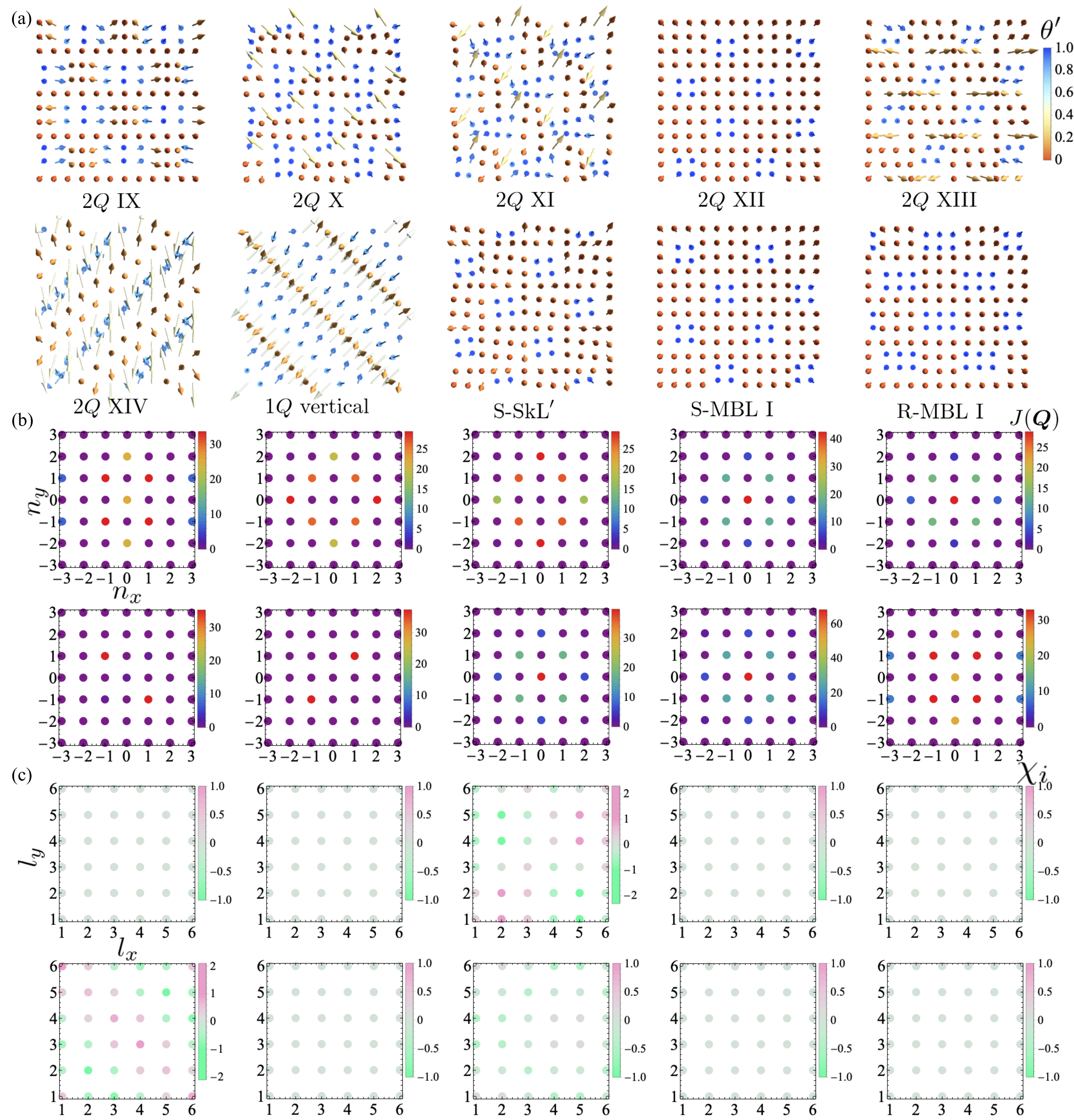}
    \caption{
    Real-space and momentum-space characteristics of the magnetic phases stabilized only in the negative-$\Delta$ region. 
    Panels (a)--(c) show (a) the three-dimensional magnetic-moment configurations in a $6 \times 6$ unit cell, with normalized moment lengths and a color scale indicating the normalized polar angle $\theta' = \theta / \pi$; 
    (b) the corresponding structure-factor distributions $J(\bm{q})$ in momentum space, with $n_x$ and $n_y$ representing multiples of $2\pi / 6$ in the first Brillouin zone ($-3 \le n_x, n_y \le 3$); 
    and (c) the local scalar chirality $\chi_i$ in the $6 \times 6$ unit cell.
    }
    \label{fig_RealspaceSfChirality_alpha_minus}
\end{figure*}

Twenty types of magnetic phases, each with its real-space magnetic-moment configuration,
except for the fully polarized state (Ferro) in the high-field region, are presented in Figs.~\ref{fig_RealspaceSfChirality_alpha_plus_78}(a), \ref{fig_RealspaceSfChirality_alpha_minus}(a), and \ref{fig_RealspaceSfChirality_previous}(a). 
The data in Figs.~\ref{fig_RealspaceSfChirality_alpha_plus_78}~and~\ref{fig_RealspaceSfChirality_previous} includes 
the states stabilized either only in the positive-$\Delta$ region or in both the positive- and negative-$\Delta$ regions,
whereas the 
data in Fig.~\ref{fig_RealspaceSfChirality_alpha_minus} includes the states stabilized only in the negative-$\Delta$ region.
For better 
visualization, the $6 \times 6$ square lattice is plotted as a $12 \times 12$ lattice by replicating the original $6 \times 6$ unit cell. 
Because the local magnetic-moment length $\langle \hat{\bm{J}}_i \rangle$ for each site differs from the spatially averaged length and the magnitude of the shift varies with the magnetic field and crystal-field splitting, we normalize the moments uniformly only for plotting the three-dimensional configurations in the panel (a)~\footnote{An example is shown in Ref.~\cite{ZhaHayamiPhysRevB.111.165155} ($\alpha$ is fixed at $0.38$): At $\Delta = 0$ and $\Delta = 4$, the magnetic-moment length $\abs{\langle \hat{\bm{J}}_{i} \rangle}$ varies from $2.47$ (2$Q$~I) to $2.50$ (Ferro) and $\abs{\langle \hat{\bm{J}}_{i} \rangle}$ varies from $2.27$ (2$Q$~III) to $2.49$ (Ferro), respectively, with magnetic field increasing.}. 
In Figs.~\ref{fig_RealspaceSfChirality_alpha_plus_78}(a)~and~\ref{fig_RealspaceSfChirality_alpha_minus}(a), we adopt a normalized polar angle $\theta'$ to represent the orientation of magnetic moments along the $z$ direction. 
Here, $\theta'$ is defined as the polar angle $\theta$ divided by $\pi$, taking values from $0$ (north pole) to $1$ (south pole) on the unit sphere in the plots.

To identify and characterize the obtained magnetic phases in the phase diagrams in Fig.~\ref{fig_phasedia_alpha}, we evaluate and present:
(i) the structure-factor distributions $J(\bm{q})$ within the first Brillouin zone in Figs.~\ref{fig_RealspaceSfChirality_alpha_plus_78}(b)~and~\ref{fig_RealspaceSfChirality_alpha_minus}(b), and the nonzero direction-resolved components, $J^{xy}_{\bm{q}}\equiv J^{x}(\bm{q})+J^{y}(\bm{q})$ and $J^{z}_{\bm{q}}\equiv J^{z}(\bm{q})$, at dominant (higher-harmonic) modes $\bm{Q}_\eta$ ($\bm{Q}'_\eta$) in Table~\ref{table:ssf_alpha},
(ii) the real-space distribution of the local scalar chirality $\chi_i$ in Figs.~\ref{fig_RealspaceSfChirality_alpha_plus_78}(c)~and~\ref{fig_RealspaceSfChirality_alpha_minus}(c)~\footnote{When plotting the local scalar chirality distribution, the colorbar range is fixed to $[-1,1]$ if the maximum value of $\chi_{i}$ is smaller than $1$, and the minimum value of $\chi_{i}$ is greater than $-1$.
Otherwise, the colorbar limits are set to the actual maximum and minimum values of $\chi_{i}$.} and its spatial average (net scalar chirality) $\langle\chi\rangle$,
(iii) the discrete topological skyrmion number $N_{\mathrm{sk}}$ (and the associated local skyrmion density $\Omega_i$),
and (iv) the net magnetization $\langle\bm{J}\rangle$ (and its magnitude).
The combined information from these complementary analyses provides a comprehensive basis for identifying the magnetic phases.
We also summarize the presence/absence of local/net scalar chirality and/or a nonzero topological skyrmion number in the ``feature'' column in Table~\ref{table:ssf_alpha}.
Note that $\chi_i = 0$ for all sites is consistent with $\abs{\langle\chi\rangle}=0$, whereas $\abs{\langle\chi\rangle}=0$ does not necessarily imply $\chi_i=0$ locally due to possible cancellation upon averaging; for the phases discussed here, $\chi_i = 0$ is also consistent with $\abs{N_{\mathrm{sk}}}=0$.
All definitions and conventions follow Ref.~\cite{ZhaHayamiPhysRevB.111.165155} (Secs.~III\,B and IV; Eqs.~(19)--(24) therein); for convenience and completeness, we summarize them in Appendix~\ref{app:physical quantities}.

\begin{table*}[htb!]
\centering
\caption{Nonzero components of $
J_{\bm{Q}_\eta}$ ($\bm{Q}_\eta \parallel [110]$) and $
J_{\bm{Q}'_\eta}$ ($\bm{Q}'_\eta \parallel [100]$) ($\eta=1,2$) in each phase, corresponding to Figs.~\ref{fig_RealspaceSfChirality_alpha_plus_78} and \ref{fig_RealspaceSfChirality_alpha_minus} in this section, and Fig.~\ref{fig_RealspaceSfChirality_previous} in Appendix~\ref{app:states}.
The ``feature'' column indicates the presence of finite local/net scalar chirality and/or a nonzero topological skyrmion number, 
while $\alpha$ and $\Delta$ represent the superposition coefficient and the crystal-field splitting, where each state typically appears.
Note that $\chi_i = 0$ for all sites is consistent with $\abs{\langle\chi\rangle}=0$, whereas $\abs{\langle\chi\rangle}=0$ does not necessarily imply $\chi_i=0$ locally due to possible cancellation upon averaging; for the phases discussed here, $\chi_i = 0$ is also consistent with $\abs{N_{\mathrm{sk}}}=0$.}
\label{table:ssf_alpha}
\vspace{2mm}
\renewcommand{\arraystretch}{1.2}
\begin{tabular}{lcccccccccccccccccc}
\hline
\hline
phase 
& $J^{xy}_{\bm{Q}_1},J^{xy}_{\bm{Q}_2}$
& $J^{z}_{\bm{Q}_1}, J^{z}_{\bm{Q}_2}$ 
& $J^{xy}_{\bm{Q}'_1}, J^{xy}_{\bm{Q}'_2}$  
& $J^{z}_{\bm{Q}'_1}, J^{z}_{\bm{Q}'_2}$
& feature 
& $\alpha$ 
& $\Delta$ \\ \hline

1$Q$~conical 
& $J^{xy}_{\bm{Q}_1}$ 
& -- 
& -- 
& -- 
& $\chi_i=0$  
& all
& $\pm$\\  

S-SkL 
&$J^{xy}_{\bm{Q}_1}=J^{xy}_{\bm{Q}_2}$
&$J^{z}_{\bm{Q}_1}=J^{z}_{\bm{Q}_2}$ 
&$J^{xy}_{\bm{Q}'_1}=J^{xy}_{\bm{Q}'_2}$
&$J^{z}_{\bm{Q}'_1}=J^{z}_{\bm{Q}'_2}$ 
&$\abs{ \langle \chi \rangle } \ne 0$, $\abs{N_{\text{sk}}}=1$
& all 
& $\pm$
\\ 

2$Q$~I
& $J^{xy}_{\bm{Q}_1}, J^{xy}_{\bm{Q}_2}$
&$J^{z}_{\bm{Q}_1}, J^{z}_{\bm{Q}_2}$ 
& $J^{xy}_{\bm{Q}'_1} \simeq J^{xy}_{\bm{Q}'_2}$
&$J^{z}_{\bm{Q}'_1}, J^{z}_{\bm{Q}'_2}$ 
& $\chi_i=0$ 
& all
& $\pm$\\ 

2$Q$~II 
& $J^{xy}_{\bm{Q}_1}, J^{xy}_{\bm{Q}_2}$
&$J^{z}_{\bm{Q}_1}, J^{z}_{\bm{Q}_2}$ 
& $J^{xy}_{\bm{Q}'_1} \simeq J^{xy}_{\bm{Q}'_2}$
&$J^{z}_{\bm{Q}'_1} \simeq J^{z}_{\bm{Q}'_2}$ 
& $ \abs{ \langle \chi \rangle} \ne 0$ 
& $0.3,0.38,0.408$ 
& $+$ \\

2$Q$~III
& $J^{xy}_{\bm{Q}_1}, J^{xy}_{\bm{Q}_2}$
&$J^{z}_{\bm{Q}_1}, J^{z}_{\bm{Q}_2}$ 
& $J^{xy}_{\bm{Q}'_1}, J^{xy}_{\bm{Q}'_2}$
&$J^{z}_{\bm{Q}'_1},J^{z}_{\bm{Q}'_2}$ 
& $\chi_i \neq 0$
& all 
& $+$ \\ 

2$Q$~IV 
&$J^{xy}_{\bm{Q}_1},J^{xy}_{\bm{Q}_2}$
&$J^{z}_{\bm{Q}_1} \simeq J^{z}_{\bm{Q}_2}$ 
&$1 \gg J^{xy}_{\bm{Q}'_1} >J^{xy}_{\bm{Q}'_2}$
&$J^{z}_{\bm{Q}'_1}, 1\gg J^{z}_{\bm{Q}'_2}$ 
& $\chi_i \neq 0$
& all
& $\pm$\\ 

2$Q$~V 
& $J^{xy}_{\bm{Q}_1} \simeq J^{xy}_{\bm{Q}_2}$
&$J^{z}_{\bm{Q}_1} \simeq J^{z}_{\bm{Q}_2}$ 
& $1 \gg J^{xy}_{\bm{Q}'_1}$
&$J^{z}_{\bm{Q}'_1}, 1\gg J^{z}_{\bm{Q}'_2}$ 
& $\chi_i \neq 0$
& all 
& $\pm$
\\ 

2$Q$~VI 
& $J^{xy}_{\bm{Q}_1}, 1\gg J^{xy}_{\bm{Q}_2}$
&$J^{z}_{\bm{Q}_1} \simeq J^{z}_{\bm{Q}_2}$ 
&$1\gg J^{xy}_{\bm{Q}'_1} \simeq J^{xy}_{\bm{Q}'_2}$
&$J^{z}_{\bm{Q}'_1}=J^{z}_{\bm{Q}'_2}$ 
& $\chi_i \neq 0$
& all
&$+$
\\ 

2$Q$~VII 
& $J^{xy}_{\bm{Q}_1}  \simeq  J^{xy}_{\bm{Q}_2}$
&$J^{z}_{\bm{Q}_1}  \simeq J^{z}_{\bm{Q}_2}$ 
& $J^{xy}_{\bm{Q}'_1}$
& $J^{z}_{\bm{Q}'_1}$ $J^{z}_{\bm{Q}'_2}$
& $\chi_i \neq 0$
& $0.6124, 0.65$ 
& $+$\\

2$Q$~VIII 
& $J^{xy}_{\bm{Q}_1}, J^{xy}_{\bm{Q}_2}$
&$J^{z}_{\bm{Q}_1} \simeq J^{z}_{\bm{Q}_2}$ 
& $1 \gg  J^{xy}_{\bm{Q}'_1} > J^{xy}_{\bm{Q}'_2} $
&$J^{z}_{\bm{Q}'_1},J^{z}_{\bm{Q}'_2}$ 
& $\chi_i \neq 0$
& $0.408, 0.6124, 0.65$ 
& $+$\\
\hline

1$Q$~vertical
& $J^{xy}_{\bm{Q}_1}$ 
& $J^{xy}_{\bm{Q}_1}$ 
& -- 
& -- 
& $\chi_i=0$  
& $0.65$
& $-$\\  

S-SkL$'$
&$J^{xy}_{\bm{Q}_1} \simeq J^{xy}_{\bm{Q}_2}$
&$J^{z}_{\bm{Q}_1}\simeq J^{z}_{\bm{Q}_2}$ 
&$J^{xy}_{\bm{Q}'_1}\simeq J^{xy}_{\bm{Q}'_2}$
&$J^{z}_{\bm{Q}'_1} \simeq J^{z}_{\bm{Q}'_2}$ 
&$\abs{\langle \chi \rangle} \ne 0$, $\abs{N_{\text{sk}}}=1$
& 0.6124 
& $-$
\\ 

S-MBL~I
&--
& $J^{z}_{\bm{Q}_1} = J^{z}_{\bm{Q}_2}$   
&--
& $J^{z}_{\bm{Q}'_1} = J^{z}_{\bm{Q}'_2}$  
& $\chi_i=0$
& $0.3, 0.38, 0.408, 0.6124$
& $-$ \\

R-MBL~I
&--
&$J^{z}_{\bm{Q}_1} = J^{z}_{\bm{Q}_2}$   
& --
&$J^{z}_{\bm{Q}'_1}$  
& $\chi_i=0$
& $0.3, 0.38, 0.408$
& $-$ \\

2$Q$~IX  
& $J^{xy}_{\bm{Q}_1} = J^{xy}_{\bm{Q}_2}$
&$J^{z}_{\bm{Q}_1} = J^{z}_{\bm{Q}_2}$   
&$1 \gg J^{xy}_{\bm{Q}'_1}$
&$J^{z}_{\bm{Q}'_1}, 1\gg J^{z}_{\bm{Q}'_2}$  
& $\chi_i=0$
& all 
& $-$ \\ 

2$Q$~X
& $J^{xy}_{\bm{Q}_1} \simeq J^{xy}_{\bm{Q}_2}$
&$J^{z}_{\bm{Q}_1} \simeq J^{z}_{\bm{Q}_2}$   
& --
& $J^{z}_{\bm{Q}'_1}, J^{z}_{\bm{Q}'_2}$  
& $\chi_i=0$
& all  
& $-$\\

2$Q$~XI  
& $J^{xy}_{\bm{Q}_1} \simeq J^{xy}_{\bm{Q}_2}$
&$J^{z}_{\bm{Q}_1} \simeq J^{z}_{\bm{Q}_2}$
&$1 \gg J^{xy}_{\bm{Q}'_1}$
& $J^{z}_{\bm{Q}'_1}, J^{z}_{\bm{Q}'_2}$  
& $\chi_i \neq 0$
& $0.6124, 0.65$  
& $-$ \\

2$Q$~XII
&$1 \gg J^{xy}_{\bm{Q}_1}=J^{xy}_{\bm{Q}_2}$
&$J^{z}_{\bm{Q}_1}=J^{z}_{\bm{Q}_2}$ 
&--
&$J^{z}_{\bm{Q}'_1}=J^{z}_{\bm{Q}'_2}$ 
& $\chi_i=0$
& $0.6124$  
& $-$
\\

2$Q$~XIII
&$J^{xy}_{\bm{Q}_1} \simeq  J^{xy}_{\bm{Q}_2}$
&$J^{z}_{\bm{Q}_1}=J^{z}_{\bm{Q}_2}$ 
&$1 \gg J^{xy}_{\bm{Q}'_1}$
&$J^{z}_{\bm{Q}'_1},J^{z}_{\bm{Q}'_2}$ 
& $\chi_i=0$
& $0.6124$  
& $-$
\\

2$Q$~XIV
&$J^{xy}_{\bm{Q}_1}, J^{xy}_{\bm{Q}_2}$
&$J^{z}_{\bm{Q}_1}$ 
&$1 \gg J^{xy}_{\bm{Q}'_1}=J^{xy}_{\bm{Q}'_2}$
&$1 \gg J^{z}_{\bm{Q}'_1}=J^{z}_{\bm{Q}'_2}$ 
& $\chi_i \neq 0$
& $0.65$
& $-$
\\

\hline\hline
\end{tabular}
\end{table*}

Among the twenty magnetic phases obtained in this study (excluding the ``Ferro''), 
the S-SkL, 1$Q$~conical spiral (1$Q$~CS), 
2$Q$~I, 2$Q$~III, 2$Q$~IV, 2$Q$~V, 2$Q$~VI, 2$Q$~IX, and 2$Q$~X states 
are robustly stabilized for all $\alpha$ values within the calculated range of crystal-field splitting $\Delta$. 
In contrast, another kind of the S-SkL (denoted as S-SkL$'$), 1$Q$~vertical spiral (1$Q$~VS), 2$Q$~II, 2$Q$~VII, 2$Q$~VIII, 2$Q$~XI, 2$Q$~XII, 2$Q$~XIII, 2$Q$~XIV, square-shaped MBL (labeled I) (S-MBL~I), and rectangular-shaped MBL (labeled I) (R-MBL~I) states exist only for specific values of $\alpha$. 
We discuss the S-SkL state in detail and briefly describe several representative states in the following subsubsections (Secs.~\ref{subsubsec:skl}--\ref{subsubsec:others}).

\subsubsection{Square skyrmion lattice}
\label{subsubsec:skl}
The most remarkable observation in this study is the emergence of the S-SkL state in the intermediate-field region in Fig.~\ref{fig_phasedia_alpha}. 
For $\alpha = 0.3$, the S-SkL state extends across the crystal-field-splitting range of $-2.5 \lesssim \Delta \lesssim 3$. 
The S-SkL region reaches its maximum extent at $\Delta \simeq 2$ for magnetic fields in the range $1 \lesssim h \lesssim 1.25$, and tends to shrink for negative values of $\Delta$, around $1.2 \lesssim h \lesssim 1.25$.
This result suggests that the S-SkL state is favored when the ground-state Kramers doublet is $\ket{\Gamma^{(1)}_{t7\pm}}$. 
In addition, the S-SkL state is no longer stabilized in the region where $|\Delta| \gg \mathcal{J}_1$, as the model effectively reduces to a single-orbital model with magnetic anisotropy. 
This behavior is consistent with previous studies, which have shown that the interplay between short-range exchange interactions and easy-axis anisotropy alone is insufficient to stabilize the S-SkL~\cite{ZhaHayamiPhysRevB.111.165155}. 
These results indicate that the interorbital coupling inherent to the multiorbital nature plays a crucial role in stabilizing the S-SkL, whose effect becomes relatively weaker compared to that of the crystal-field splitting, $|\Delta|$, for $|\Delta| \gg \mathcal{J}_1$.

As the crystal-field splitting varies from $\Delta = 2.5$ to $-2$, the absolute value of the net scalar chirality, $\abs{\langle \chi \rangle}$, also changes significantly. 
It reaches a maximum of approximately $3.478$ at $\Delta = 2.5$ and $h = 1.2$ under the magnetic-field interval defined in Sec.~\ref{subsec:alpha}, and decreases to $0.543$ at $\Delta = -2$ and $h = 1.25$, accompanied by an enhanced easy-axis anisotropy along the $z$ direction. 
The absolute value of the topological skyrmion number, $\abs{N_{\text{sk}}}$, remains quantized at $1$. 
In the absence of bond-dependent anisotropy~\cite{hayamidoi:10.7566/JPSJ.89.103702} and/or DMI, the $N_{\text{sk}} = +1$ and $N_{\text{sk}} = -1$ states are energetically degenerate, and the skyrmion helicity is not fixed. 
The structure factor exhibits fourfold rotational symmetry, with peak positions satisfying
$J^{xy}_{\bm{Q}_1} = J^{xy}_{\bm{Q}_2}$, 
$J^{z}_{\bm{Q}_1} = J^{z}_{\bm{Q}_2}$, 
$J^{xy}_{\bm{Q}'_1} = J^{xy}_{\bm{Q}'_2}$, and 
$J^{z}_{\bm{Q}'_1} = J^{z}_{\bm{Q}'_2}$, as summarized in Table~\ref{table:ssf_alpha}.

\begin{figure}[htbp]
    \begin{center}
    \includegraphics[width=8cm]{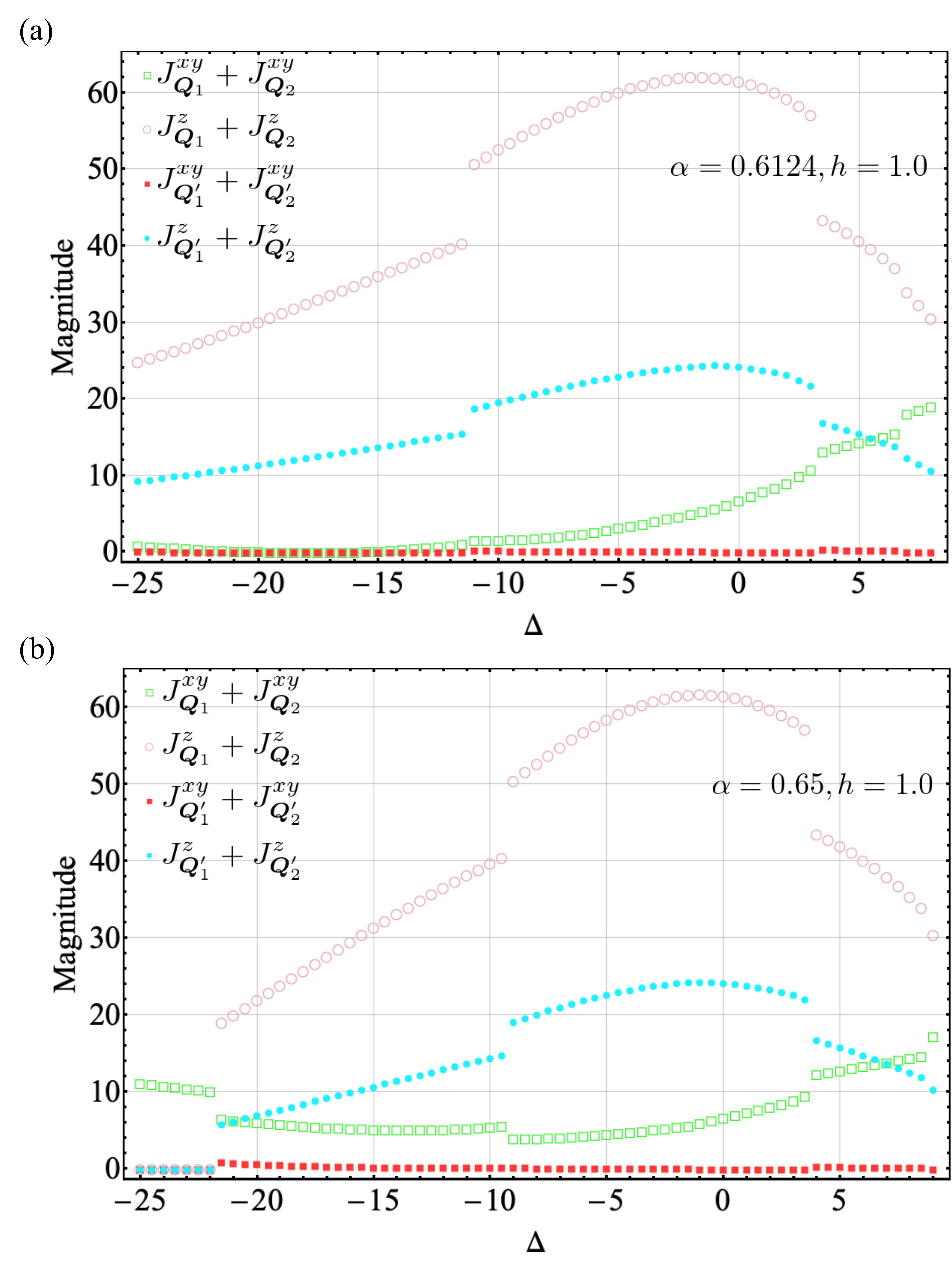}
    \caption{Magnitude of the summed structure factor components (vertical axis), $J^{xy}_{\bm{Q}_1} + J^{xy}_{\bm{Q}_2}$, $J^{z}_{\bm{Q}_1} + J^{z}_{\bm{Q}_2}$, $J^{xy}_{\bm{Q}'_1} + J^{xy}_{\bm{Q}'_2}$, and $J^{z}_{\bm{Q}'_1} + J^{z}_{\bm{Q}'_2}$, for (a) $\alpha = 0.6124$ and (b) $\alpha = 0.65$ at fixed magnetic field $h = 1$, plotted as functions of the crystal-field splitting $\Delta$ (horizontal axis).}
    \label{fig_delta_SF}
    \end{center}
\end{figure}

With increasing $\alpha$, the easy-axis anisotropy of both $\Gamma_{t7}$ Kramers doublets weakens simultaneously [Fig.~\ref{fig_alpha}(a)], while that between
$\ket{\Gamma^{(1)}_{t7}}$ and $\ket{\Gamma^{(2)}_{t7}}$
doublets strengthen.
In such a situation, the region of the S-SkL state tends to expand. 
The S-SkL state appears in the range $-3 \lesssim \Delta \lesssim 3.5$ at $\alpha = 0.38$ [Fig.~\ref{fig_phasedia_alpha}(b)] and $-3.5 \lesssim \Delta \lesssim 3.5$ at $\alpha = 0.408$ [Fig.~\ref{fig_phasedia_alpha}(c)]. 
For $\alpha = 0.6124$ and $0.65$, both $\Gamma_{t7}$ doublets exhibit easy-plane anisotropy in the $xy$ plane, whereas the interorbital anisotropy
approaches
its maximum along the easy axis.
In this situation, the phase diagrams show that the S-SkL state appears over a much wider range of crystal-field splitting $\Delta$ compared with $\alpha = 0.3$, $0.38$, and $0.408$. 
At $\alpha = 0.6124$ [Fig.~\ref{fig_phasedia_alpha}(d)], the S-SkL state is stabilized within $-6.5 \lesssim \Delta \lesssim 7.5$. 
These tendencies suggest that the interplay between the intraorbital easy-plane anisotropy and interorbital easy-axis anisotropy plays a significant role in enlarging the S-SkL region against $\Delta$.
When the crystal-field splitting varies from $\Delta = 7$ to $-6$, $\abs{\langle \chi \rangle}$ reaches a maximum of approximately $3.022$ at $\Delta = 6.5$ and $h = 1.2$, and decreases to $0.913$ at $\Delta = -6$ and $h = 1.26$. 
With decreasing $\Delta$ and $\abs{\langle \chi \rangle}$, the easy-axis anisotropy
is significantly enhanced; such a tendency is found in the weight of the direction-resolved structure factor, as shown in Fig.~\ref{fig_delta_SF}(a). 

For crystal-field splitting in the range $-25 \lesssim \Delta \lesssim -6$ at $\alpha=0.6124$,  
an emergent magnetic-moment configuration with a finite topological skyrmion number $\abs{N_{\text{sk}}} = 1$, denoted S-SkL$'$, replaces the S-SkL state.
The S-SkL$'$ exhibits stronger easy-axis anisotropy [Fig.~\ref{fig_RealspaceSfChirality_alpha_minus}(a)] and a smaller scalar chirality [Fig.~\ref{fig_RealspaceSfChirality_alpha_minus}(c)] compared with the S-SkL. 
In contrast to the S-SkL, the fourfold rotational symmetry in this structure is slightly broken, as reflected by the nearly but not exactly equal peak intensities,
$J^{xy}_{\bm{Q}_1} \simeq J^{xy}_{\bm{Q}_2}$, 
$J^{z}_{\bm{Q}_1} \simeq J^{z}_{\bm{Q}_2}$, 
$J^{xy}_{\bm{Q}'_1} \simeq J^{xy}_{\bm{Q}'_2}$, and 
$J^{z}_{\bm{Q}'_1} \simeq J^{z}_{\bm{Q}'_2}$, as shown in Table~\ref{table:ssf_alpha}.
The breaking of fourfold rotational symmetry in the S-SkL$'$ is presumably attributed to the strong easy-axis anisotropy. 
Because of this extreme anisotropy, even a slight imbalance in the magnetic-moment configuration can destabilize the fourfold rotational symmetry of the S-SkL state.
These results indicate that excessively strong easy-axis anisotropy 
suppresses the formation of the fourfold-symmetric S-SkL.

The stability region of the S-SkL is further extended in the case of $\alpha = 0.65$, as shown in Fig.~\ref{fig_phasedia_alpha}(e).
Starting from large positive $\Delta$ ($\Delta = 8.5$) and moving toward negative $\Delta$, the S-SkL region reaches its minimum width around $\Delta = -1.5$ to $-2$. 
As $\Delta$ further decreases, the S-SkL region gradually expands, which survives at least up to $\Delta=-25$. 
The scalar chirality $\abs{\langle \chi \rangle}$ decreases rapidly from $\Delta = 8.5$ to approximately $\Delta \simeq -9$. 
The easy-axis anisotropy of the S-SkL correspondingly increases up to 
$\Delta \simeq -9$ at $h=1$ [Fig.~\ref{fig_delta_SF}(b)],
similarly to the case of $\alpha = 0.6124$.
Unlike the case for $\alpha = 0.6124$, however, $\abs{\langle \chi \rangle}$ remains nearly constant within $0.9 \lesssim \abs{\langle \chi \rangle} \lesssim 1.3$ upon further decreasing $\Delta$, rather than diminishing continuously. 
At the same time, the easy-axis anisotropy of the S-SkL remains the dominant anisotropy in the system but becomes weaker than that for $\alpha = 0.6124$, as compared in Figs.~\ref{fig_delta_SF}(a) and \ref{fig_delta_SF}(b) at $h=1$.
The S-SkL state with weak but still dominant easy-axis anisotropy persists down to our lowest calculated value of $\Delta = -25$.

In addition, in the low-field region, the 1$Q$~VS state appears at $\Delta = -18$, followed by the emergence of the 2$Q$~XIV state at $\Delta = -22.5$.
This replacement of the low-field phase leads to a sequence of phase transitions against $h$,
1$Q$~VS $\rightarrow$ 2$Q$ $\rightarrow$ S-SkL $\rightarrow$ 1$Q$~CS $\rightarrow$ Ferro, which closely resembles that reported in the model incorporating the dominant ordering wave vector, higher-harmonic wave vector, easy-axis anisotropy, and Zeeman coupling~\cite{HayamiPhysRevB.105.174437}.
These findings indicate that the multiorbital model can reproduce the results of the single-orbital spin-only model by appropriately taking into account the crystal-field splitting and magnetic anisotropy.

\begin{figure}[htbp]
    \begin{center}
    \includegraphics[width=8cm]{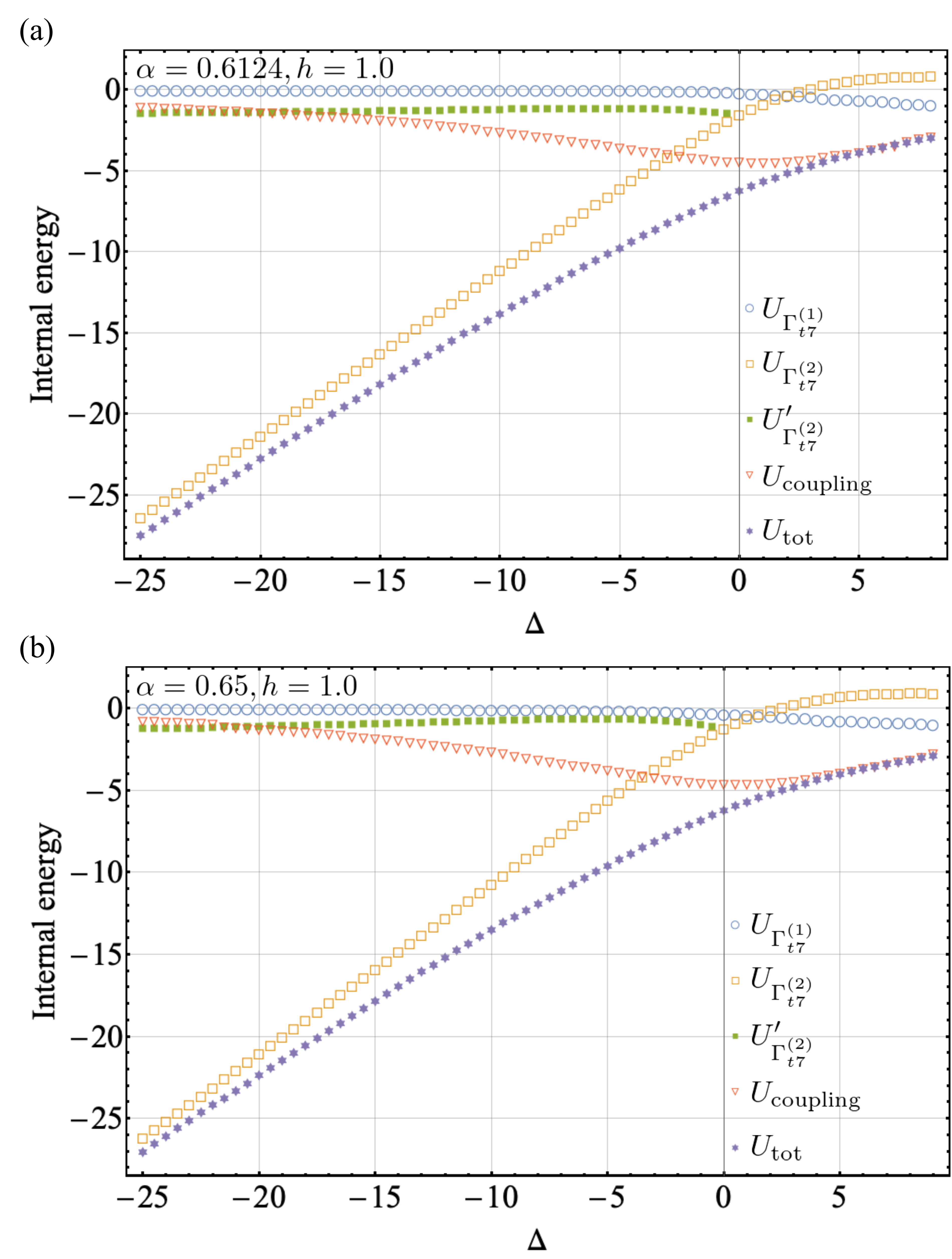}
    \caption{Decomposed and total internal energies $U_{\Gamma^{(1)}_{t7}}$, $U_{\Gamma^{(2)}_{t7}}$, $U_{\text{coupling}}$, and $U_{\text{tot}}$ as a function
    of the crystal-field splitting $\Delta$ for (a) $\alpha = 0.6124$ and (b) $\alpha = 0.65$ at $h = 1$.
    The shifted energy $U'_{\Gamma^{(2)}_{t7}}$ is also plotted in the negative-$\Delta$ regime to compare the energy order between $U'_{\Gamma^{(2)}_{t7}}$ and $U_{\text{coupling}}$.}
    \label{fig_delta_energy}
    \end{center}
\end{figure}

To elucidate the contribution of each diagonal and off-diagonal block to the energy
gain, 
we decompose the total effective Hamiltonian $\hat{\mathcal{H}}_{\text{tot}}$ in \Eq{eq:TotMFhamiltonian} as
\begin{align}\label{eq:Htot_decomposed}
\hat{\mathcal{H}}_{\text{tot}}
= \hat{\mathcal H}_{\Gamma^{(1)}_{t7}}
+ \hat{\mathcal H}_{\Gamma^{(2)}_{t7}}
+ \hat{\mathcal H}_{\text{coupling}},
\end{align}
where $\hat{\mathcal H}_{\Gamma^{(1)}_{t7}}$ and $\hat{\mathcal H}_{\Gamma^{(2)}_{t7}}$ correspond to the top-left and bottom-right
$2\times2$ subspaces, respectively, while $\hat{\mathcal H}_{\text{coupling}}$ represents the off-diagonal $2\times2$ blocks.
With the thermal average $\langle\cdot\rangle$ defined in Eq.~(\ref{eq:exactDia}),
the internal energy $U$ is expressed as
\begin{align}\label{eq:Utot_decomposed}
U_{\text{tot}} \equiv \langle \hat{\mathcal H}_{\text{tot}} \rangle
&= \langle \hat{\mathcal H}_{\Gamma^{(1)}_{t7}}\rangle
+ \langle \hat{\mathcal H}_{\Gamma^{(2)}_{t7}} \rangle
+ \langle \hat{\mathcal H}_{\text{coupling}} \rangle \nonumber \\
&= U_{\Gamma^{(1)}_{t7}}
+ U_{\Gamma^{(2)}_{t7}}
+ U_{\text{coupling}},
\end{align}
where the expectation values are evaluated in the eigenstate basis $\{ \ket{n} \}$ of $\hat{\mathcal{H}}_{\text{tot}}$.
Figures~\ref{fig_delta_energy}(a) and \ref{fig_delta_energy}(b) show the decomposed and total internal energies as functions of $\Delta$ 
for $\alpha = 0.6124$ and $0.65$, respectively, under $h = 1$, corresponding to the S-SkL state~\footnote{For large positive crystal-field splitting ($\Delta \gtrsim 7.5$ for $\alpha = 0.6124$, $\Delta \gtrsim 9$ for $\alpha = 0.65$), the S-SkL state is replaced by the 2$Q$~VI phase as shown in Figs.~\ref{fig_phasedia_alpha}(d)--\ref{fig_phasedia_alpha}(e). 
In contrast, for large negative splitting ($\Delta \lesssim -22$), the S-SkL state is replaced by the 1$Q$~CS phase.}.
A crossover occurs around $\Delta \simeq 2$ for $\alpha = 0.6124$ and $\Delta \simeq 1.5$ for $\alpha = 0.65$: 
above these values, the $\ket{\Gamma^{(1)}_{t7\pm}}$ doublet predominantly lowers the energy, 
whereas below them, the $\ket{\Gamma^{(2)}_{t7\pm}}$ doublet becomes dominant.
For $\Delta < 0$, the negative splitting further stabilizes the $\ket{\Gamma^{(2)}_{t7\pm}}$ doublet, 
leading to an almost linear decrease in $U_{\Gamma^{(2)}_{t7}}$ with decreasing $\Delta$.
To remove the trivial linear offset induced by $\Delta$, 
we define the shifted energies as 
$U'_{\Gamma^{(2)}_{t7}} = U_{\Gamma^{(2)}_{t7}} - \Delta$ 
and 
$U'_{\Gamma^{(1)}_{t7}} = U_{\Gamma^{(1)}_{t7}} + \Delta$, 
which facilitates comparison of their relative contributions of exchange interactions.
To highlight the behavior of the $\ket{\Gamma^{(2)}_{t7}}$ doublet, we also plot $U'_{\Gamma^{(2)}_{t7}}$ in the negative-$\Delta$ regime ($-25 \le \Delta \le -0.5$) in Fig.~\ref{fig_delta_energy}.
Even for a large negative $\Delta$, 
the interorbital coupling term $U_{\text{coupling}}$ remains finite, 
indicating that the multiorbital character is not completely quenched. 
The shifted energy $U'_{\Gamma^{(2)}_{t7}}$ is of the same order as $U_{\text{coupling}}$ (both being approximately unity), indicating that the interorbital
coupling remains energetically relevant even for large negative $\Delta$.
The large negative $\Delta$ also enhances the easy-axis anisotropy of the S-SkL state at $\alpha = 0.6124$ 
by strengthening the effective single-ion anisotropy within each orbital channel.
These results confirm the crucial role of multiorbital effects in stabilizing the S-SkL state.
Since it is difficult to directly evaluate the respective contributions of higher-harmonic wave vectors and intraorbital coupling induced by $\Delta$ to the total energy reduction, 
the above discussion is valid only for $\alpha \neq 0.65$. 
For $\alpha = 0.65$, based on numerical results and comparison with previous studies, 
we empirically attribute the enhanced occurrence of the S-SkL states
to the competition between the momentum-dependent frustrated exchange interactions and the intraorbital coupling characterized by $\Delta$. 
The former effectively weakens the easy-axis anisotropy and thereby promotes the stabilization of additional S-SkL states.

\subsubsection{2$Q$~II state}
\label{subsubsec:2QII}
\begin{figure}[htbp]
    \begin{center}
    \includegraphics[width=7.5cm]{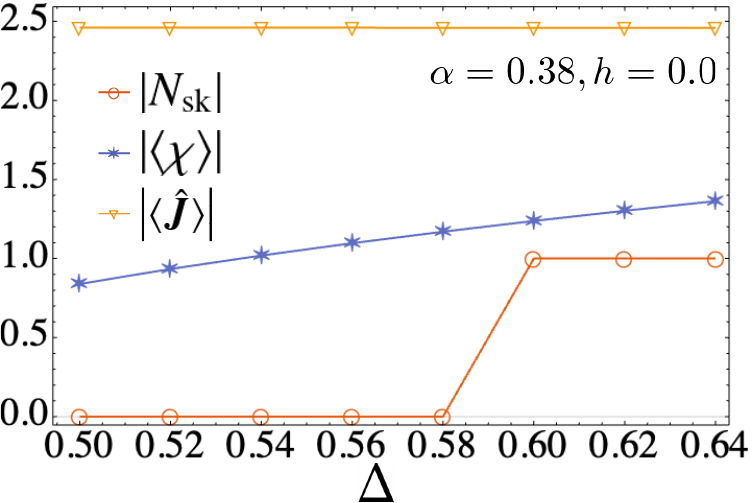}
        \caption{Dependence of the absolute skyrmion number $\abs{N_{\text{sk}}}$, the absolute net scalar chirality $\abs{\langle \chi \rangle}$, and the magnetization magnitude $\abs{\langle \hat{\bm{J}}\rangle}$
        on the crystal-field splitting $\Delta$ for $\alpha = 0.38$ and $h = 0$.}
    \label{fig_fluct}
    \end{center}
\end{figure}

The 2$Q$~II state is another double-$Q$ phase characterized by a finite net scalar chirality, 
as summarized in Table~\ref{table:ssf_alpha}.
In the present study, the 2$Q$~II state appears within a narrow range of $\Delta \simeq 0.5$ for $\alpha = 0.3$, $0.38$, and $0.408$ in the low-field region, and its stability range further shrinks with increasing $\alpha$. 
Our previous work Ref.~\cite{ZhaHayamiPhysRevB.111.165155} reported in
detail that this phase is stabilized in the region $0.5 \lesssim \Delta \lesssim 0.64$ and $0 \leq h \lesssim 0.15$ at $\alpha = 0.38$. 
In addition, the skyrmion number of this state depends sensitively on $\Delta$: it is zero for $0.5 \lesssim \Delta \lesssim 0.6$, while it becomes finite for $0.6 \lesssim \Delta \lesssim 0.64$. 
Reference~\cite{ZhaHayamiPhysRevB.111.165155} also pointed out that the fluctuation in the skyrmion number may originate from the nearly coplanar magnetic-moment configuration of the 2$Q$~II state. 
Figure~\ref{fig_fluct} shows the $\Delta$ dependence of the absolute skyrmion number $\abs{N_{\text{sk}}}$, the absolute net scalar chirality $\abs{\langle \chi \rangle}$, and the magnetization magnitude $\abs{\langle \hat{\bm{J}}\rangle}$, for $\alpha = 0.38$ and $h = 0 $, obtained from the data in Ref.~\cite{ZhaHayamiPhysRevB.111.165155}. 
Moreover, the similarity in the structure-factor distributions among the 2$Q$~I, 2$Q$~II, and 2$Q$~III states suggests that the 2$Q$~II state acts as an intermediate state connecting the 2$Q$~I and 2$Q$~III states.

\subsubsection{2$Q$~IV state}
\label{subsubsec:2QIV}
The 2$Q$~IV state is characterized by comparable but different intensities of $J^{xy}_{\bm{Q}_1}$ and $J^{xy}_{\bm{Q}_2}$, together with small but finite components at the higher-harmonic wave vectors
$\bm{Q}'_1$ and $\bm{Q}'_2$.
The out-of-plane components also appear
with nearly the same intensities at the dominant ordering wave vectors, indicating a
noncoplanar magnetic-moment configuration with broken fourfold rotational symmetry 
as summarized in Table~\ref{table:ssf_alpha}.
The net scalar chirality $\langle \chi \rangle$ vanishes, confirming its topologically trivial
nature. 
This phase predominantly appears in the positive-$\Delta$ region and begins to adjoin the S-SkL state from $\Delta \simeq 1$ for $\alpha = 0.3$, $0.38$, and $0.408$, from $\Delta \simeq 2.5$ for $\alpha = 0.6124$, and from $\Delta \simeq 3$ for $\alpha = 0.65$, in an intermediate field region 
(Fig.~\ref{fig_phasedia_alpha}).
These results indicate a strong competition between the 2$Q$~IV and S-SkL states.

\subsubsection{2$Q$~V and 2$Q$~IX states}
\label{subsubsec:2QVandIX}
The 2$Q$~V state is stabilized in the low- and intermediate-field regions for both positive and negative $\Delta$ values as shown in Fig.~\ref{fig_phasedia_alpha}. 
It typically appears near the boundary between the 2$Q$~IX state in the negative-$\Delta$ region and the 2$Q$~IV state in the positive-$\Delta$ region, as well as near the S-SkL state on both sides of $\Delta$ for all $\alpha$ in the intermediate-field region. 
Its nearly equal amplitudes of $J^{xy}_{\bm{Q}_1}$ and $J^{xy}_{\bm{Q}_2}$, together with comparable $J^{z}_{\bm{Q}_1}$ and $J^{z}_{\bm{Q}_2}$, indicate a balanced double-$Q$ superposition. 
Meanwhile, the presence of $J^{z}_{\bm{Q}'_1}$ and the strongly suppressed $J^{xy}_{\bm{Q}'_1}$ and $J^{z}_{\bm{Q}'_2}$ components suggest that anisotropic higher-harmonic distortions exist. 
This configuration corresponds to a noncoplanar but topologically trivial magnetic texture without the net scalar chirality.

In contrast, the 2$Q$~IX state emerges on the negative-$\Delta$ side for all $\alpha$
(Fig.~\ref{fig_phasedia_alpha}). 
It preserves equal-amplitude dominant double-$Q$ modulations in both the in-plane and out-of-plane components at $\bm{Q}_1$ and $\bm{Q}_2$ but exhibits unequal finite higher-harmonic terms as the 
2$Q$~V state.
The 2$Q$~IX state is an almost coplanar magnetic-moment configuration. 
The absence of finite local scalar chirality ($\chi_i = 0$) identifies it as a nontopological, easy-axis double-$Q$ phase. 
We regard the 2$Q$~IX state as the outcome of a phase transition from the 
2$Q$~V state that occurs upon decreasing $\Delta$.

\subsubsection{Square and rectangular magnetic bubble lattice~I states}
\label{subsubsec:MBLs}
A MBL is a two-dimensional nontopological soliton bubble lattice composed of multi-$Q$ ordering waves. 
In contrast to the SkL state, 
the MBL state is characterized by a zero topological skyrmion number and a vanishing local/net scalar chirality~\cite{HayamiPhysRevB.93.184413,hayamibubblePhysRevB.108.024426}. 
To distinguish the MBL states from other topologically trivial multi-$Q$ states discussed in this work (e.g., 2$Q$~I/IX), we emphasize that the MBLs are characterized by a longitudinal modulation: the transverse components vanish ($J^{xy}_{\bm{Q}_\eta}=J^{xy}_{\bm{Q}'_\eta}=0$), while characteristic higher-harmonic peaks appear in the longitudinal component (see also Table~\ref{table:ssf_alpha}).
The checkerboard-type S-MBL~I phase corresponds to a two-dimensional soliton lattice consisting of periodically aligned magnetic-moment bubbles with a vanishing topological charge. 
Its structure factors exhibit identical peaks at $\bm{Q}_1$ and $\bm{Q}_2$, together with their equal higher-harmonic components $\bm{Q}'_1$ and $\bm{Q}'_2$, reflecting a fourfold rotational symmetry.

The R-MBL~I phase is another nontopological soliton lattice but with broken fourfold rotational symmetry. 
Compared with the S-MBL~I phase, its structure factors exhibit identical peaks at $\bm{Q}_1$ and $\bm{Q}_2$, while only the $J^{z}_{\bm{Q}'_1}$ component remains as a higher-harmonic term, leading to the symmetry breaking. 
The local/net scalar chirality remains zero, consistent with its nontopological nature. 

In particular, the S-MBL~I state can be regarded as the result of a phase transition from the topologically nontrivial S-SkL (or S-SkL$'$) state that occurs upon increasing $\abs{\Delta}$ in the negative-$\Delta$ region [Figs.~\ref{fig_phasedia_alpha}(a)--\ref{fig_phasedia_alpha}(d)]. 
It can also be regarded as the outcome of a phase transition from the R-MBL~I state as the magnetic field increases from the low-field to the intermediate-field region [Figs.~\ref{fig_phasedia_alpha}(a)--\ref{fig_phasedia_alpha}(c)].

\subsubsection{2$Q$~XII state}
\label{subsubsec:2QXII}
The 2$Q$~XII state emerges between the S-SkL$'$ and S-MBL~I states at $\alpha = 0.6124$, as shown in Fig.~\ref{fig_phasedia_alpha}(d). 
Its structure factors contain finite but small and equal components of $J^{xy}_{\bm{Q}_1}$ and $J^{xy}_{\bm{Q}_2}$, together with $J^{z}_{\bm{Q}_{1}} =J^{z}_{\bm{Q}_{2}}$ and the higher-harmonic terms $J^{z}_{\bm{Q}'_{1}} = J^{z}_{\bm{Q}'_{2}}$. 
These features imply a partial deformation of the S-SkL$'$ state without complete bubble formation of the S-MBL~I state (Fig.~\ref{fig_RealspaceSfChirality_alpha_minus}).
This phase represents the onset of the crossover from the S-SkL$'$ to the S-MBL~I state, where the S-SkL$'$ state gradually loses its 
scalar chirality while maintaining a double-$Q$ modulation.

\subsubsection{Other states}
\label{subsubsec:others}
The remaining magnetic phases are not discussed in detail here; however, their characteristics are summarized in Table~\ref{table:ssf_alpha} and illustrated in Figs.~\ref{fig_RealspaceSfChirality_alpha_plus_78}, \ref{fig_RealspaceSfChirality_alpha_minus}
and
\ref{fig_RealspaceSfChirality_previous}.

\subsection{Effects of interorbital coupling}
\label{subsec:weighting}
In this subsection, to examine the role of the interorbital
coupling on the stability of the S-SkL, we introduce a weighting coefficient $\gamma$ that multiplies the $2 \times 2$ top-right and bottom-left off-diagonal blocks in the $4 \times 4$ total angular-momentum operator to control the strength of the interorbital coupling, where $\gamma$ ranges from $0$ to $1$. 
We decompose the original total angular-momentum operator $\hat{\bm{J}}_i$ into a linear combination of $\hat{\bm{J}}_i^{\text{diag}}$, which contains only the diagonal block, and $\hat{\bm{J}}_i^{\text{off-diag}}$, which contains only the off-diagonal block in $\hat{\mathcal{H}}_{\text{ex}}$:  
\begin{align}
    \hat{\bm{J}}_i = \hat{\bm{J}}_i^{\text{diag}} + \gamma\, \hat{\bm{J}}_i^{\text{off-diag}}, \quad \gamma \in [0, 1].
\end{align}
In the limit of $\gamma = 0$, the crystal-field ground-state orbitals at each site are completely decoupled, meaning that interorbital processes are fully separated from the intraorbital 
processes.
It should be emphasized that, when evaluating the converged expectation value $\langle \hat{\bm{J}}_j \rangle$ used for calculating physical quantities, we always employ the original operator as the quantum-mechanical 
observable ($\gamma = 1$). 
Likewise, for the Zeeman coupling, since the magnetic field couples directly to the local magnetic moment at each site, the original operator ($\gamma = 1$) must also be used.

\begin{figure}[htbp]
    \begin{center}
    \includegraphics[width=8.0cm]{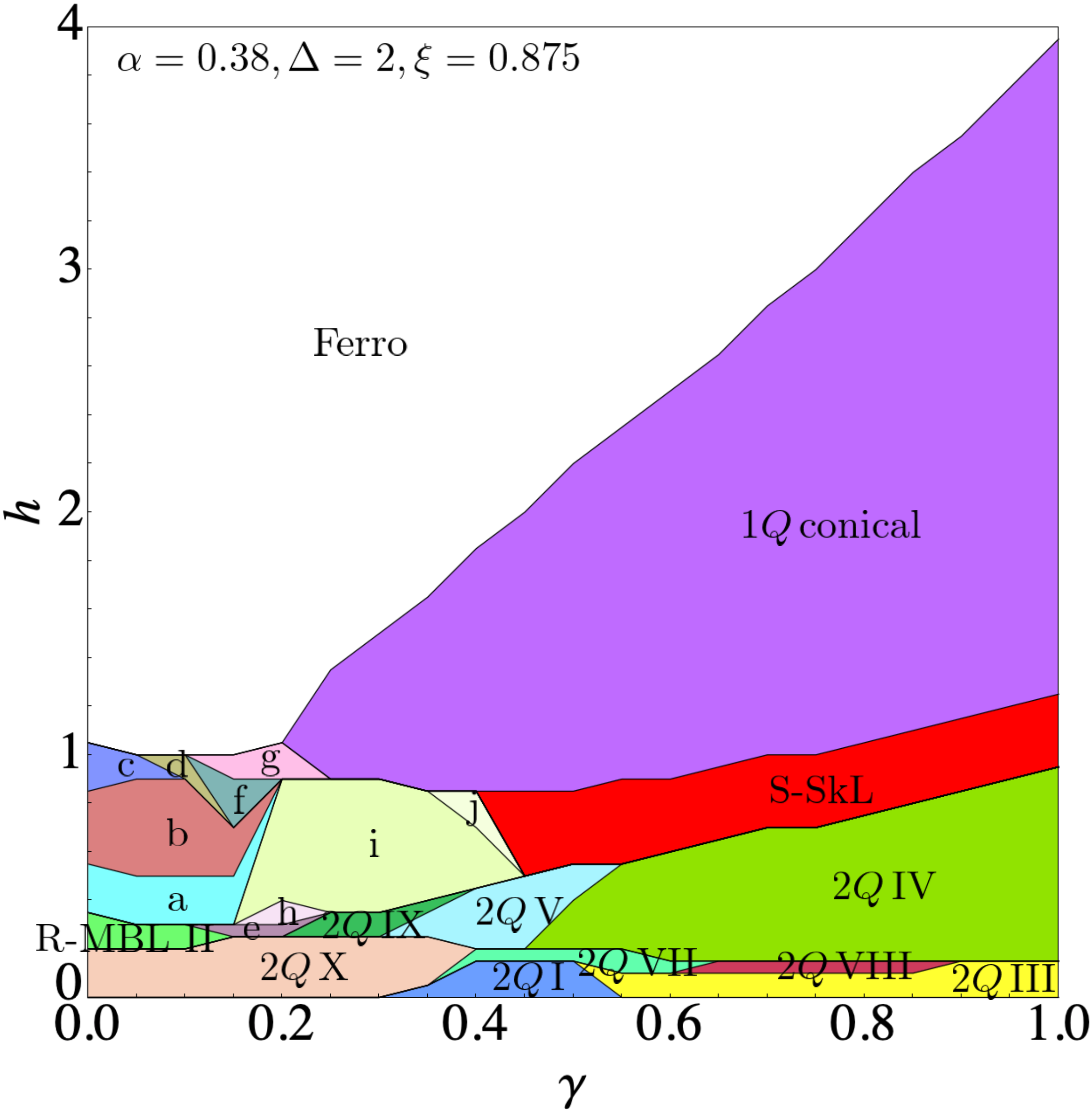}
    \caption{
    Phase diagram at low temperature ($T = 0.05$) as functions of the magnetic field $h$ (vertical axis) and the weighting coefficient $\gamma$ (horizontal axis). 
    Each colored region corresponds to a distinct low-temperature magnetic-moment configuration, including the 1$Q$~CS state, various previously observed 2$Q$ states (labeled I, III--V, and VII--X), 
    emergent 2$Q$ states (labeled a--j, where the prefix ``2$Q$'' is omitted for simplicity), the S-SkL state, the R-MBL~II state (which differs from R-MBL~I in its real-space configuration as shown in Fig.~\ref{fig_RealspaceSfChirality_offdia}(a)), and the fully polarized ferromagnetic state (Ferro).}
    \label{fig_phasedia_offdia}
    \end{center}
\end{figure}

\begin{figure*}[t!]
    \begin{center}
    \includegraphics[width=1.0 \hsize]{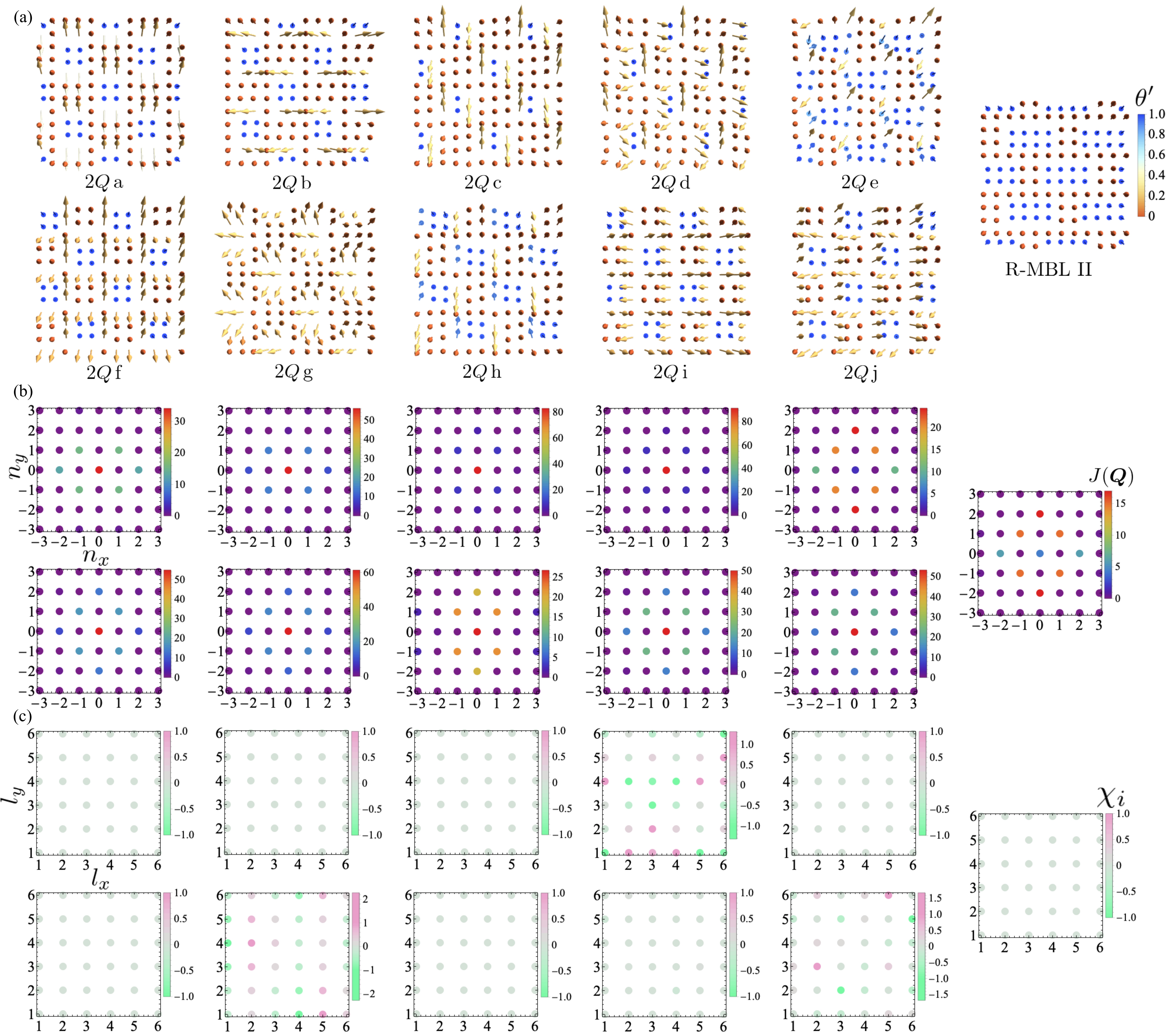}
    \caption{
    Real-space and momentum-space characteristics of magnetic-moment configurations obtained from various values of the weighting coefficient $\gamma$.
    Panels (a) and (b) display three aspects of the real-space and momentum-space physical quantities:
    (a)
    Three-dimensional magnetic-moment configurations in a $6 \times 6$ unit cell, with magnetic moments drawn at each site and normalized in length for clarity. 
    The color scale encodes the normalized polar angle $\theta' = \theta / \pi$, varying continuously from 0 (north pole) to 1 (south pole).
    (b)
    Structure factor
    distributions $
    J(\bm{q})$ regarding the magnetic moments in
    momentum space. 
    The axis labels $n_{x}$ and $n_{y}$ denote the multiples of ${2\pi}/6$ with $-3 \leq n_{x}, n_{y} \leq 3$ in the first Brillouin zone.
    (c)
    Local scalar chirality $\chi_i$ in the $6 \times 6$ unit cell.}
    \label{fig_RealspaceSfChirality_offdia}
    \end{center}
\end{figure*}

Figure~\ref{fig_phasedia_offdia} shows the $\gamma$--$h$ phase diagrams for $\alpha = 0.38$ and $\Delta = 2$ at a low temperature of $T = 0.05$ and a higher-harmonic wave-vector parameter of $\xi = 0.875$. 
The S-SkL state exists in the range of $0.4 \lesssim \gamma \le 1$, which indicates the importance of the contribution from the interorbital coupling.
Notably, the emergence of the 2$Q$~VII phase, which does not appear at $\alpha = 0.3$, $0.38$, or $0.408$, and the 2$Q$~VIII phase, which is absent at $\alpha = 0.3$ and $0.38$, indicates that, at $\alpha = 0.38$, the system exhibits a trend similar to that for $\alpha = 0.6124$ and $0.65$. 
The weaker interorbital coupling leads to enhanced orbital selectivity and effectively influences the anisotropy between the diagonal and off-diagonal orbital sectors, thereby promoting the emergence of 
double-$Q$ states.
Notably, we find another topologically trivial soliton state as the interorbital coupling is weakened, R-MBL~II, which does not appear in the $\alpha$-dependent phase diagrams in Sec.~\ref{subsec:alpha}. 
In contrast to R-MBL~I, R-MBL~II has a finite $J^z_{\bm{Q}'_2}$ component, leading to a distinct real-space magnetic texture.
Table~\ref{table:ssf_offdia} and Fig.~\ref{fig_RealspaceSfChirality_offdia} summarize the characteristics of the 
obtained
magnetic-moment configurations. 

These results confirm that the interorbital coupling (i.e., multiorbital effect) plays a crucial role in stabilizing the S-SkL state in $f$-electron multiorbital systems.

\begin{table*}[htb!]
\centering
\caption{Nonzero components of $
J_{\bm{Q}_\eta}$ ($\bm{Q}_\eta \parallel [110]$) and $
J_{\bm{Q}'_\eta}$ ($\bm{Q}'_\eta \parallel [100]$) ($\eta=1,2$) in each phase, corresponding to Fig.~\ref{fig_phasedia_offdia}. 
\label{table:ssf_offdia}}
\vspace{2mm}
\renewcommand{\arraystretch}{1.2}
\begin{tabular}{lcccccccccccccccccc}
\hline
\hline
phase 
& $J^{xy}_{\bm{Q}_1}, J^{xy}_{\bm{Q}_2}$
& $J^{z}_{\bm{Q}_1}, J^{z}_{\bm{Q}_2}$ 
& $J^{xy}_{\bm{Q}'_1}, J^{xy}_{\bm{Q}'_2}$  
& $J^{z}_{\bm{Q}'_1}, J^{z}_{\bm{Q}'_2}$
& feature  
\\ 
\hline

R-MBL~II
& --
& $J^{z}_{\bm{Q}_1} = J^{z}_{\bm{Q}_2}$
& -- 
& $J^{z}_{\bm{Q}'_1}, J^{z}_{\bm{Q}'_2}$
& $\chi_i=0$   
\\

2$Q$~a
& $J^{xy}_{\bm{Q}_1} = J^{xy}_{\bm{Q}_2}$
& $J^{z}_{\bm{Q}_1} = J^{z}_{\bm{Q}_2}$
& $ J^{xy}_{\bm{Q}'_1} , J^{xy}_{\bm{Q}'_2}$
& $J^{z}_{\bm{Q}'_1} , J^{z}_{\bm{Q}'_2}$
& $\chi_i=0$   
\\

2$Q$~b
& $J^{xy}_{\bm{Q}_1} = J^{xy}_{\bm{Q}_2}$
& $J^{z}_{\bm{Q}_1} = J^{z}_{\bm{Q}_2}$
& $ J^{xy}_{\bm{Q}'_1}$
& $J^{z}_{\bm{Q}'_1} , J^{z}_{\bm{Q}'_2}$
& $\chi_i=0$   
\\

2$Q$~c
& $J^{xy}_{\bm{Q}_1} = J^{xy}_{\bm{Q}_2}$
& $J^{z}_{\bm{Q}_1} = J^{z}_{\bm{Q}_2}$
& $ J^{xy}_{\bm{Q}'_1}$
& $J^{z}_{\bm{Q}'_1} , J^{z}_{\bm{Q}'_2}$
& $\chi_i=0$   
\\

2$Q$~d
& $J^{xy}_{\bm{Q}_1} = J^{xy}_{\bm{Q}_2}$
& $J^{z}_{\bm{Q}_1} = J^{z}_{\bm{Q}_2}$
& $J^{xy}_{\bm{Q}'_1}, J^{xy}_{\bm{Q}'_2}$
& $J^{z}_{\bm{Q}'_1}, J^{z}_{\bm{Q}'_2}$
& $\chi_i \neq 0$ 
\\

2$Q$~e
& $J^{xy}_{\bm{Q}_1} = J^{xy}_{\bm{Q}_2}$
& $J^{z}_{\bm{Q}_1} = J^{z}_{\bm{Q}_2}$
& $ J^{xy}_{\bm{Q}'_1}$
& $J^{z}_{\bm{Q}'_1} , J^{z}_{\bm{Q}'_2}$
& $\chi_i=0$  
 \\

2$Q$~f
& $J^{xy}_{\bm{Q}_1} = J^{xy}_{\bm{Q}_2}$
& $J^{z}_{\bm{Q}_1} = J^{z}_{\bm{Q}_2}$
& $J^{xy}_{\bm{Q}'_1}, J^{xy}_{\bm{Q}'_2}$
& $J^{z}_{\bm{Q}'_1}, J^{z}_{\bm{Q}'_2}$
& $\chi_i=0$  
\\

2$Q$~g
& $J^{xy}_{\bm{Q}_1} \simeq J^{xy}_{\bm{Q}_2}$
& $J^{z}_{\bm{Q}_1} \simeq J^{z}_{\bm{Q}_2}$
& $J^{xy}_{\bm{Q}'_1}, J^{xy}_{\bm{Q}'_2}$
& $J^{z}_{\bm{Q}'_1}, J^{z}_{\bm{Q}'_2}$
& $\chi_i \neq 0$
\\

2$Q$~h
& $J^{xy}_{\bm{Q}_1} \simeq J^{xy}_{\bm{Q}_2}$
& $J^{z}_{\bm{Q}_1} \simeq J^{z}_{\bm{Q}_2}$
& $J^{xy}_{\bm{Q}'_1}, J^{xy}_{\bm{Q}'_2}$
& $J^{z}_{\bm{Q}'_1}, J^{z}_{\bm{Q}'_2}$
& $\chi_i=0$  
\\

2$Q$~i
& --
& $J^{z}_{\bm{Q}_1} = J^{z}_{\bm{Q}_2}$
& $J^{xy}_{\bm{Q}'_1} \simeq J^{xy}_{\bm{Q}'_2}$
& $J^{z}_{\bm{Q}'_1} \simeq J^{z}_{\bm{Q}'_2}$
& $\chi_i=0$  
\\

2$Q$~j
& $J^{xy}_{\bm{Q}_1}$
& $J^{z}_{\bm{Q}_1} \simeq J^{z}_{\bm{Q}_2}$
& $J^{xy}_{\bm{Q}'_1} \simeq J^{xy}_{\bm{Q}'_2}$
& $J^{z}_{\bm{Q}'_1}\simeq J^{z}_{\bm{Q}'_2}$
& $\chi_i \neq 0$ 
\\

\hline\hline
\end{tabular}
\end{table*}

\subsection{Effects of interactions at higher-harmonic wave vectors}
\label{subsec:higherharmonics}
A number of previous studies have reported that higher-harmonic wave vectors play a crucial role in stabilizing the S-SkL, the MBLs, and other trivial double-$Q$ states on the square lattice in centrosymmetric frustrated systems without DMI~\cite{ZTWangPhysRevB.103.104408,Hayamidoi:10.7566/JPSJ.91.023705,HayamiPhysRevB.105.174437,HAYAMI2023170547,Hayamimagnetism5020012}. 
In this subsection, we further explore
the role
of the higher-harmonic wave vectors
in the $f$-electron multiorbital model.
The existence of a double-$Q$ superposition of helical spirals at $\bm{Q}_1$ and $\bm{Q}_2$, i.e., the S-SkL, naturally leads to the intensities at the higher-harmonic wave vectors at $\bm{Q}'_1$ and $\bm{Q}'_2$.
Meanwhile, the superposition of double-$Q$ sinusoidals at $\bm{Q}_1$ and $\bm{Q}_2$ generally does not lead to the intensities at $\bm{Q}'_1$ and $\bm{Q}'_2$.
Thus, the interactions at $\bm{Q}'_1$ and $\bm{Q}'_2$ affect the type of 
double-$Q$ superposition.
Their magnitude is evaluated by the Fourier components of the real-space interaction, where $\xi = 0.875$ is used in previous results~\cite{ZhaHayamiPhysRevB.111.165155}.

\begin{figure}[htbp]
    \begin{center}
    \includegraphics[width=8.0cm]{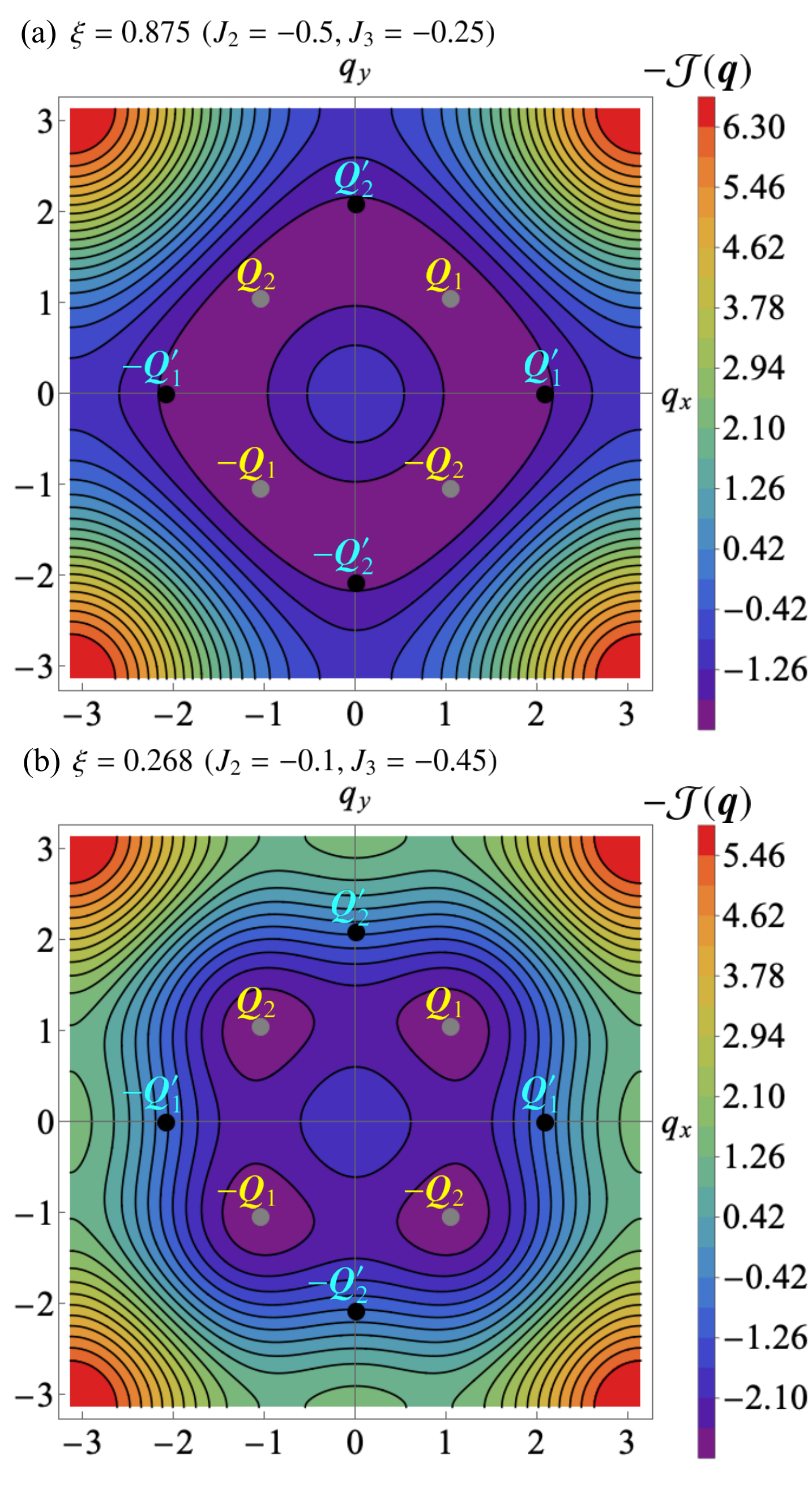}
    \caption{(a) Contour maps of the negative Fourier-transformed exchange interaction $-\mathcal{J}(\bm{q})$ for $\xi=0.875$ ($\mathcal{J}_2=-0.5$, $\mathcal{J}_3=-0.25$) and (b) $\xi=0.268$ ($\mathcal{J}_2=-0.1$, $\mathcal{J}_3=-0.45$). The color scale represents the value of $\mathcal{J}(\bm{q})$ (red: larger, blue: smaller). The dominant ordering vectors are $\bm{Q}_1=(2\pi/6,\,2\pi/6)$ and $\bm{Q}_2=(-2\pi/6,\,2\pi/6)$ (and their negatives), and the higher-harmonic vectors are defined by $\bm{Q}'_{1}\equiv \bm{Q}_1-\bm{Q}_2$ and $\bm{Q}'_{2}\equiv \bm{Q}_1+\bm{Q}_2$.}
    \label{fig_JQdispersion}
    \end{center}
\end{figure}

Here, we consider a regime with weak higher-harmonic contributions by setting
$\xi = 0.268$ (corresponding to $\mathcal{J}_2 = -0.1$ and $\mathcal{J}_3 = -0.45$) while keeping the positions of the ordering wave vectors.
Figure~\ref{fig_JQdispersion} provides an intuitive momentum-space view of the negative Fourier-transformed exchange interaction $-\mathcal{J}(\bm{q})$ for the two representative higher-harmonic regimes.
In both panels, the dominant ordering vectors $\pm\bm{Q}_{1,2}$ are located at degenerate extrema of $-\mathcal{J}(\bm{q})$, which set the primary instability channel.
The key difference appears at the higher-harmonic vectors $\pm\bm{Q}'_{1,2}$: the local curvatures of $-\mathcal{J}(\bm{q})$ around $\bm{Q}_{\eta}$ and $\bm{Q}'_{\eta}$ characterize the stiffness of the momentum-space landscape near the dominant and higher-harmonic modes, respectively.
Consequently, in the low-$\xi$ regime [Fig.~\ref{fig_JQdispersion}(b)], the higher-harmonic extrema are less competitive (much higher in height) and typically sharper than the dominant minima, which increases the energetic penalty for developing higher-harmonic components.
By contrast, in the high-$\xi$ regime [Fig.~\ref{fig_JQdispersion}(a)], the higher harmonics become more competitive (much closer in height) with the dominant wave vectors and the landscape around $\bm{Q}'_{\eta}$ is effectively softer, facilitating momentum-space frustration between $\bm{Q}_{\eta}$ and $\bm{Q}'_{\eta}$.  

\begin{figure}[htbp]
    \begin{center}
    \includegraphics[width=8.0cm]{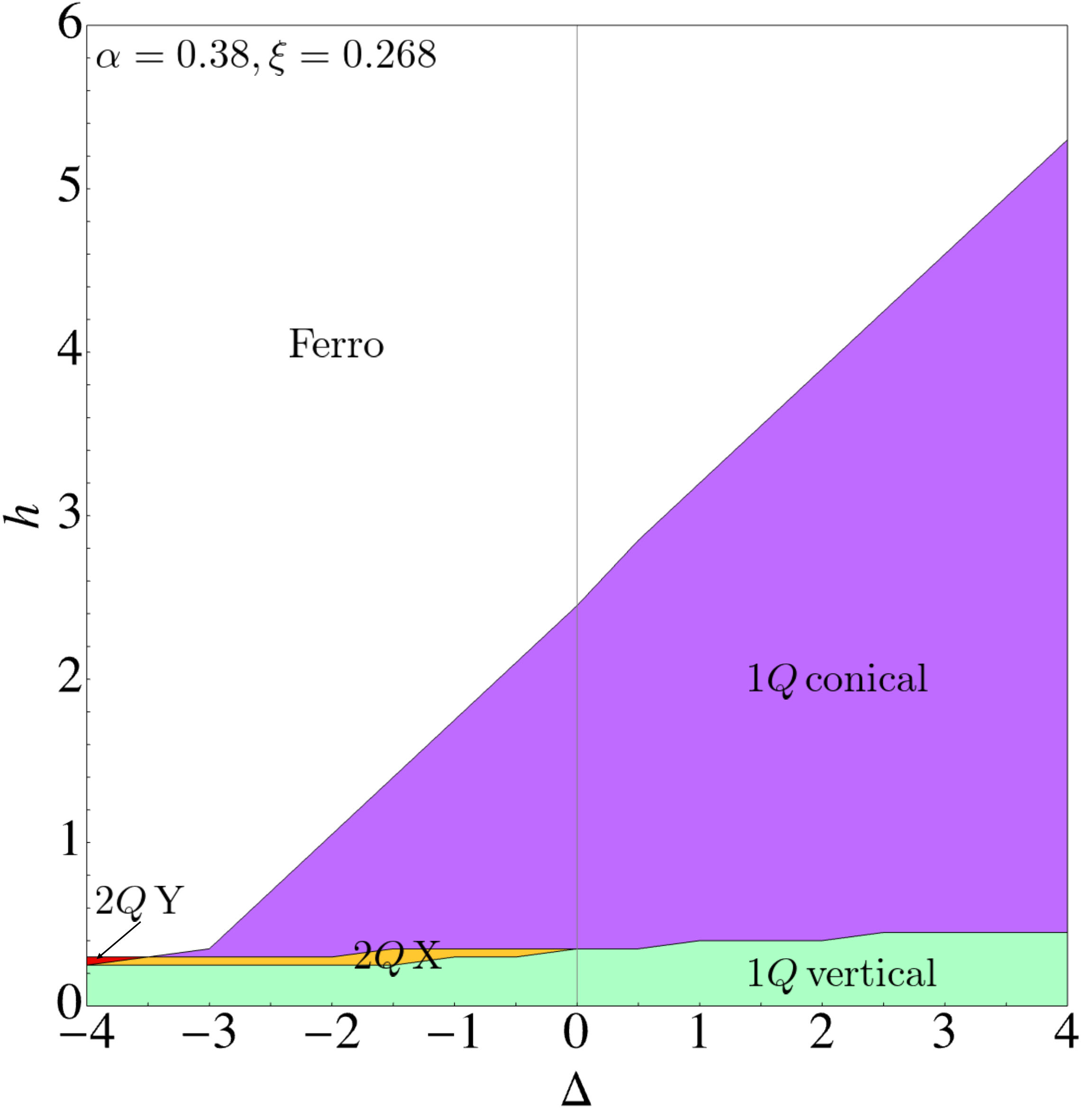}
    \caption{Phase diagram at low temperature ($T=0.05$) as functions of the magnetic field $h$ (vertical axis) and the
    crystal-field splitting $\Delta$ between the two Kramers
    doublets (horizontal axis) for $\alpha=0.38$ and $\xi= 0.268$. Each colored region corresponds to a different low-temperature magnetic-moment configuration, including the two 1$Q$ states, two 2$Q$ states (labeled X and Y)
    and the
    fully polarized ferromagnetic state (Ferro).}
    \label{fig_phasedia_without_higherhamonicWV}
    \end{center}
\end{figure}

\begin{figure}[htbp]
    \begin{center}
    \includegraphics[width=8.0cm]{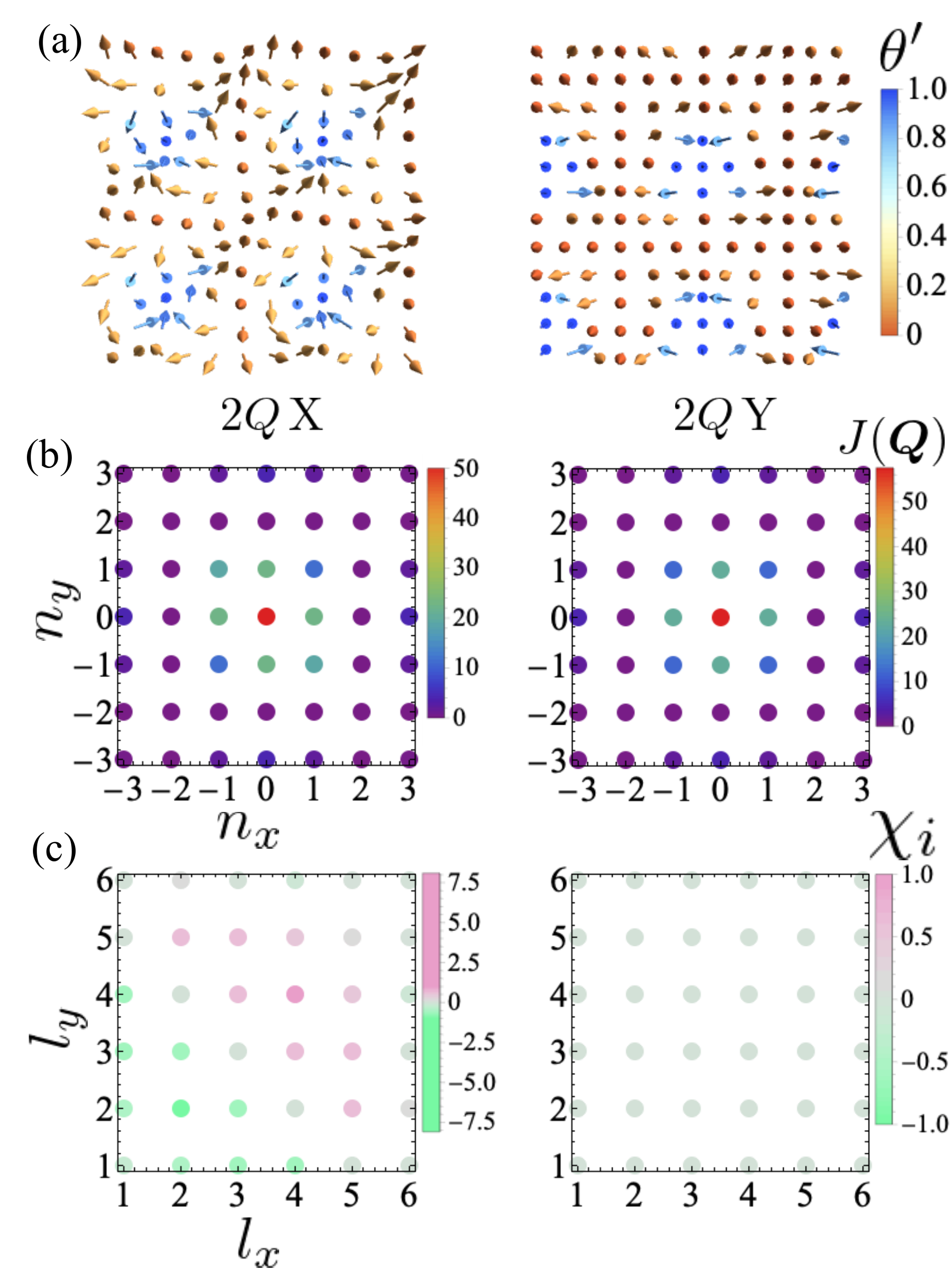}
    \caption{
    Real-space and momentum-space characteristics of magnetic-moment configurations obtained at $\alpha = 0.38$ and $\xi = 0.268$.
    Panels (a)--(c) display three aspects of the real-space and momentum-space physical quantities:
    (a)
    Three-dimensional magnetic-moment configurations in a $6 \times 6$ unit cell, with magnetic moments drawn at each site and normalized in length for clarity. 
    The color scale encodes the normalized polar angle $\theta' = \theta / \pi$, varying continuously from 0 (north pole) to 1 (south pole).
    (b)
    Structure factor
    distributions $
    J(\bm{q})$ regarding the magnetic moments in
    momentum space. 
    The axis labels $n_{x}$ and $n_{y}$ denote the multiples of ${2\pi}/6$ with $-3 \leq n_{x}, n_{y} \leq 3$ in the first Brillouin zone.
    (c)
    Local scalar
    chirality $\chi_{i}$
    in the $6 \times 6$
    unit cell.
    }
    \label{fig_RealspaceSfChirality_higherharmonic}
    \end{center}
\end{figure}

Figure~\ref{fig_phasedia_without_higherhamonicWV} shows the $\Delta$--$h$ phase diagram for $\alpha = 0.38$ at a low temperature of $T = 0.05$ and a higher-harmonic wave-vector parameter of $\xi = 0.268$. 
In this case, no topologically nontrivial magnetic-moment configurations such as the S-SkL are found, and no abundant double-$Q$ states appear.
Table~\ref{table:ssf_higherhamonicWV} and Fig.~\ref{fig_RealspaceSfChirality_higherharmonic} summarize the characteristics of the 
obtained magnetic-moment configurations. 
Specifically, only the 1$Q$~VS, 1$Q$~CS, and two 
emergent 2$Q$ states (labeled X and Y) appear. 
As shown in Fig.~\ref{fig_RealspaceSfChirality_higherharmonic}(b), the $\bm{Q}_1$ and $\bm{Q}_2$ components in these two 2$Q$ states are composed of lower-harmonic wave vectors expressed as $(\bm{Q}_1 \pm \bm{Q}_2)/2$ rather than $(\bm{Q}_1 \pm \bm{Q}_2)$.
These qualitative differences in the phase diagrams originate from the $\xi$-dependent contour of $\mathcal{J}(\bm{q})$, which
indicates that, even in the presence of multiorbital effects, the higher-harmonic wave vectors still play a crucial role in the emergence of the S-SkL state.

\begin{table*}[htb!]
\centering
\caption{Nonzero components of $
J_{\bm{Q}_\eta}$ ($\bm{Q}_\eta \parallel [110]$) and $
J_{\bm{Q}'_\eta}$ ($\bm{Q}'_\eta \parallel [100]$) ($\eta=1,2$) in each phase, corresponding to Fig.~\ref{fig_phasedia_without_higherhamonicWV}. 
\label{table:ssf_higherhamonicWV}}
\vspace{2mm}
\renewcommand{\arraystretch}{1.2}
\begin{tabular}{lcccccccccccccccccc}
\hline
\hline
phase 
& $J^{xy}_{\bm{Q}_1}, J^{xy}_{\bm{Q}_2}$
& $J^{z}_{\bm{Q}_1}, J^{z}_{\bm{Q}_2}$ 
& $J^{xy}_{\bm{Q}'_1}, J^{xy}_{\bm{Q}'_2}$  
& $J^{z}_{\bm{Q}'_1}, J^{z}_{\bm{Q}'_2}$
& feature 
& $\Delta$ \\ \hline

2$Q$~X
&$J^{xy}_{\bm{Q}_1},J^{xy}_{\bm{Q}_2}$
&$J^{z}_{\bm{Q}_1} \simeq J^{z}_{\bm{Q}_2}$ 
&$1\gg J^{xy}_{\bm{Q}'_1}=J^{xy}_{\bm{Q}'_2}$
&$1\gg J^{z}_{\bm{Q}'_1}=J^{z}_{\bm{Q}'_2}$ 
& $\chi_i \neq 0$
& $-$
\\ 

2$Q$~Y
&$J^{xy}_{\bm{Q}_1}=J^{xy}_{\bm{Q}_2}$
&$J^{z}_{\bm{Q}_1}=J^{z}_{\bm{Q}_2}$ 
&--
&$1\gg J^{z}_{\bm{Q}'_1}=J^{z}_{\bm{Q}'_2}$ 
& $\chi_i=0$ 
& $-$
\\ 

\hline\hline
\end{tabular}
\end{table*}

\section{Discussion}
\label{sec:discussion}
In Ref.~\cite{ZhaHayamiPhysRevB.111.165155}, we established the minimal ingredients required to stabilize the S-SkL in centrosymmetric tetragonal $4f$-electron multiorbital systems. 
The present work goes beyond that baseline not only by extending the parameter regimes, but more importantly by providing mechanism-resolving diagnostics and controlled tests of 
the three microscopic parameters $\alpha$, $\gamma$, and 
$\xi$, as outlined in Sec.~\ref{sec:introduction}. Below we discuss the resulting 
physical insights (Sec.~\ref{subsec:key}), the emergent soliton 
states (Sec.~\ref{subsec:emergent}), and experimental 
implications (Sec.~\ref{subsec:experimental}).

For $f$-electron magnets with unquenched orbital angular momentum, we also briefly discuss the possibility of multipolar (tensorial) exchange beyond the bilinear form and relegate technical details to Appendix~\ref{app:tensor_exchange}.

\subsection{Key microscopic parameters}\label{subsec:key}
Ref.~\cite{ZhaHayamiPhysRevB.111.165155} mainly discussed parameter sets with fixed $\alpha$ at $0.38$, whereas here we comprehensively explore the \emph{matrix-element-resolved anisotropy} over an extended $\alpha$ regime (Sec.~\ref{subsec:alpha}).
Specifically, we analyze the relevant diagonal and off-diagonal matrix elements of $\hat{\bm{J}}$ in the two-Kramers-doublet subspace as functions of the superposition coefficient $\alpha$ [Eqs.~(\ref{eq:a1}) and (\ref{eq:a2}); Fig.~\ref{fig_alpha}; Table~\ref{table:alpha_anisotropy}]. 
This provides a microscopic interpretation of the competition between intraorbital anisotropy and interorbital coupling and clarifies why the S-SkL stability window becomes significantly broader around $\alpha\simeq 0.6$.
In particular, Fig.~\ref{fig_delta_SF} quantifies how the relative weights of the structure-factor components at $\bm{Q}_\eta$ and $\bm{Q}'_\eta$ evolve with $\Delta$ at $\alpha=0.6124$ and $0.65$, providing an at-a-glance indicator of the anisotropy balance that accompanies the robust S-SkL region in these cases.

To further elucidate the broader S-SkL stability window,
we introduce an \emph{energy-resolved mechanism and orbital-sector decomposition}.
We decompose $\hat{\mathcal{H}}_{\text{tot}}$ and the internal energy into the two crystal-field-orbital sectors and their coupling channel [Eqs.~(\ref{eq:Htot_decomposed}) and (\ref{eq:Utot_decomposed}); Fig.~\ref{fig_delta_energy}]. 
This analysis identifies the energetic origin and stability window of the S-SkL beyond a parameter scan and shows that the interorbital-coupling channel remains energetically relevant even in the large-$|\Delta|$ regime.

To demonstrate the significance of the interorbital coupling, we perform a \emph{controlled test of interorbital coupling strength}
by introducing a weighting coefficient $\gamma$ to tune the off-diagonal blocks (interorbital channel) and mapping out $\gamma$--$h$ phase diagrams (Sec.~\ref{subsec:weighting}). 
This controlled procedure provides direct evidence that the S-SkL exists only within a finite window of the interorbital coupling strength and reveals additional double-$Q$ phases, including an 
emergent bubble state R-MBL~II, 
which
appears as the interorbital channel is weakened.

We also quantify the higher-harmonic effect by the dimensionless ratio $\xi = \mathcal{J}_{\bm{Q}'_{\eta}}/\mathcal{J}_{\bm{Q}_\eta}$ [Eq.~(\ref{eq:xi})] and explicitly compare strong- and weak-higher-harmonic regimes to demonstrate the \emph{momentum-based frustrated exchange and higher-harmonic selection}
(Sec.~\ref{subsec:higherharmonics}). 
The $\bm{q}$--$\mathcal{J}(\bm{q})$ contour maps (Fig.~\ref{fig_JQdispersion}) provide an intuitive momentum-space landscape: a larger $\xi$ makes the higher-harmonic modes more competitive with the dominant modes, facilitating momentum-space frustration between $\bm{Q}_\eta$ and $\bm{Q}'_\eta$ and thereby promoting the S-SkL and related multi-$Q$ instabilities, whereas a smaller $\xi$ suppresses such higher-harmonic components and eliminates the S-SkL.

\subsection{Emergent soliton states}
\label{subsec:emergent}
Beyond cataloging multi-$Q$ textures, the present work identifies an additional topologically nontrivial S-SkL state (S-SkL$'$) with a slight breaking of fourfold rotational symmetry, which differs from the ordinary S-SkL, and $\abs{N_{\text{sk} }}= 1$ (Sec.~\ref{subsubsec:skl}; Table~\ref{table:ssf_alpha}).
We also identify square and rectangular MBLs as distinct nontopological soliton (bubble) lattices and characterize them by symmetry and higher-harmonic fingerprints (Secs.~\ref{subsubsec:MBLs} and~\ref{subsec:weighting}; Tables~\ref{table:ssf_alpha} and~\ref{table:ssf_offdia}). 
This classification clarifies their physically meaningful distinction from other topologically trivial multi-$Q$ states and highlights S-SkL$'$ and MBLs as an important outcome of the multiorbital/higher-harmonic mechanism.

\subsection{Experimental implications}
\label{subsec:experimental}
Since momentum-dependent frustrated exchange interactions can select finite-$q$ ordering wave vectors in momentum space~\cite{OkuboPhysRevLett.108.017206}, momentum-resolved probes such as magnetic resonant x-ray scattering (RXS) and small-angle neutron scattering (SANS) could detect magnetic Bragg peaks at the dominant $\bm{Q}_{\eta}$ and, when higher-harmonic components are appreciable, at $\bm{Q}'_{\eta}=\bm{Q}_1\pm\bm{Q}_2$ and their symmetry-equivalent positions~\cite{khanh2020nanometric,takagi2022square}.

For crystal-field effects, several experimental techniques have been demonstrated to be powerful tools in previous studies of tetragonal Ce-based heavy-fermion materials Ce$T$In$_5$ ($T$ = Co, Rh, Ir), as introduced by Sundermann \textit{et al}.~\cite{SundermannPhysRevB.99.235143}.
Inelastic neutron scattering has been employed to determine the crystal-field splitting in Ce$T$In$_5$~\cite{ADChristiansonPhysRevB.66.193102,ADChristiansonPhysRevB.70.134505}, 
while linearly polarized soft x-ray absorption spectroscopy (XAS) has enabled detailed investigations of the ground-state wave functions~\cite{TWillersPhysRevB.81.195114,ThomasWillersdoi:10.1073/pnas.1415657112,KChenPhysRevB.97.045134}.

\section{SUMMARY} 
\label{sec:summary}
In summary, we have theoretically established a cooperative mechanism for the stabilization of the 
topologically nontrivial states
in prototypical centrosymmetric tetragonal Ce-based $4f$-electron systems, where the multiorbital effects and momentum-dependent frustrated exchange interactions---including higher-harmonic wave vectors---act simultaneously. 
Starting from two $\Gamma_{t7}$ Kramers doublets of Ce$^{3+}$, the total angular-momentum operator consists of anisotropic diagonal blocks and interorbital off-diagonal blocks. 
Self-consistent mean-field calculations on a $6\times6$ square lattice reveal that 
(i) the S-SkL is stabilized in an intermediate magnetic-field region across broad ranges of the crystal-field splitting $\Delta$ and the superposition coefficient $\alpha$, while an intraorbital excessively large $|\Delta|$ enhances the easy-axis anisotropy; 
(ii) energy decomposition confirms finite contributions from both doublets and a persistent interorbital-coupling channel even at relatively large negative $\Delta$, underscoring the essential role of multiorbital effects; 
(iii) introducing a weighting factor on the off-diagonal blocks further demonstrates that interorbital coupling governs the emergence and competition of multiple
2$Q$ phases surrounding the S-SkL; and 
(iv) suppressing the higher-harmonic contributions (small $\xi$) eliminates S-SkL formation and markedly reduces the diversity of double-$Q$ states. 

Overall, our findings identify the S-SkL as a robust consequence of the cooperative interplay between crystal-field-induced multiorbital effects and momentum-dependent frustrated exchange interactions, thereby broadening the design principles for skyrmion-hosting materials beyond Gd/Eu-based systems with quenched orbital moments.
Incorporating collective excitations and fluctuation effects (going beyond the static mean-field framework) would provide further insight into the stability of the competing multi-$Q$ phases; this important direction
will be left for future studies.

We finally comment on the possible realization of the $f$-electron multiorbital S-SkL in real materials. 
In Ce-based centrosymmetric tetragonal compounds, the crystal-field splitting $\Delta$ typically exceeds the nearest-neighbor exchange interaction $\mathcal{J}_1$ by one to several tens of times.  
For example, in CeRhIn$_5$, the crystal-field excitations at about $6.9$ meV and $23.6$ meV~\cite{ADChristiansonPhysRevB.66.193102} coexist with an in-plane nearest-neighbor exchange $\mathcal{J}_1\approx 0.74$ meV~\cite{DasPhysRevLett.113.246403}, yielding $\Delta/\mathcal{J}_1\sim 10\text{--}32$.
Thus $\Delta\approx20\text{--}30\times \mathcal{J}_1$ is a rough empirical scale for Ce-based $f$-electron systems. 
In this study, we 
found the emergence of the 
topologically nontrivial magnetic phase in cases where the crystal-field splitting satisfies $\abs{\Delta} > 10 \mathcal{J}_1$ at $\alpha = 0.6124$ and $0.65$, assuming $\mathcal{J}_1 = 1$ as the energy unit. For other cases ($\alpha = 0.3, 0.38$ and $0.408$), the S-SkL state only exists in the low-splitting region.
We therefore suggest that experimental efforts focus on searching for Ce-based centrosymmetric tetragonal magnets with intermediate $\alpha$, where intraorbital easy-plane anisotropy is enhanced, and the S-SkL phase is significantly extended, and
relatively weak crystal-field splitting. 

More broadly, our results highlight a general cooperative mechanism between multiorbital effects and momentum-based frustrated exchange that promotes multi-$Q$ instabilities via higher-harmonic wave vectors. 
This mechanism is not tied to a specific material: it requires only (i) at least two low-lying Kramers doublets with competing intraorbital and interorbital anisotropies, and (ii) a frustrated exchange interaction landscape in which the higher-harmonic ratio 
$\xi = \mathcal{J}_{\bm{Q}'_{\eta}}/\mathcal{J}_{\bm{Q}_\eta}$ is appreciable,
and may therefore provide useful guidance beyond tetragonal Ce-based compounds, e.g., to other centrosymmetric lattices and unquenched-orbital-angular-momentum $f$-electron systems, where analogous competing wave vectors and interorbital coupling can be realized.
We note that Araki \textit{et al}. reported a topological Hall effect in PrSb$_2$~\cite{araki_doi:10.7566/JPSJ.94.113703}; however, the microscopic origin and the underlying magnetic texture remain unresolved, and alternative interpretations cannot be excluded at present.

Beyond the S-SkL in centrosymmetric $4f$-electron systems, we note that Seo \textit{et al}. reported a topologically trivial MBL state in the tetragonal Ce-based compound CeAuSb$_2$~\cite{seo2021spin}. 
This material therefore provides a promising platform to explore transport signatures with nontopological multi-$Q$ magnetic textures and to further test the general interplay between competing ordering wave vectors and multiorbital coupling discussed in the present work.

\begin{acknowledgments}

Y.S.Z. would like to express his gratitude to T. Shirato and T. Yamanaka from Hokkaido University for
their fruitful discussions. Y.S.Z. is also grateful to X.-R. Hou, an alumnus of the University of Science and Technology of China, for his valuable comments on data processing.
This research was supported by JSPS KAKENHI Grants Numbers JP22H00101, JP22H01183, JP23H04869, JP23K03288, and by JST CREST (JPMJCR23O4) and JST FOREST (JPMJFR2366).
Y.S.Z. was supported by JST SPRING (Grant Number JPMJSP2119), which funded the presentation of this work.

\end{acknowledgments}

\appendix
\section{CRYSTAL-FIELD HAMILTONIAN}
\label{app:CEF}
The electric multipoles contribute to the crystal-field Hamiltonian, which is written as
\begin{align}
    \hat{\mathcal{H}}_{\text{CEF}} = B_{2}^{0}\hat{O}_{2}^{0}
    + B_{4}^{0}\hat{O}_{4}^{0} + B_{4}^{4}\hat{O}_{4}^{4},
\end{align}    
where $B_{2}^{0}$, $B_{4}^{0}$, and $B_{4}^{4}$ are the crystal-field parameters.
$\hat{O}_{2}^{0}$
and ($\hat{O}_{4}^{0}$, $\hat{O}_{4}^{4}$) represent one rank-2 electric quadrupole and two rank-4 electric hexadecapoles, respectively. 
These multipoles $\hat{O}_{k}^{q}$ are expressed as polynomials of the total angular-momentum operators, known as Stevens operators~\cite{stevens1952matrix,hutchings1964point}, defined by
\begin{eqnarray}
\label{eq:stevens}
\hat{O}_{2}^{0} &=& 3(\hat{J}^z)^{2} - \hat{\bm{J}}^{2},  \nonumber\\
\hat{O}_{4}^{0} &=& 35(\hat{J}^z)^{4} - 30\hat{\bm{J}}^{2}(\hat{J}^z)^{2} + 25(\hat{J}^z)^{2} - 6\hat{\bm{J}}^{2} + 3\hat{\bm{J}}^{4}, \nonumber\\
\hat{O}_{4}^{4} &=& \frac{1}{2} \left[ (\hat{J}^{+})^4 + (\hat{J}^{-})^4 \right]. 
\end{eqnarray}

By restricting the crystal-field Hamiltonian to the $J=5/2$ sextet (six-dimensional) subspace and diagonalizing the resulting $6\times6$ matrix, we obtain three Kramers doublets. 
Each doublet is expressed as a superposition of the $\ket{5/2,J^z}$ basis states, with the superposition coefficients determined by the crystal-field parameters.
In particular, for the two $\Gamma_{t7}$ doublets, the mixing coefficient $\alpha$ and  $\beta$ can be written as
\begin{align}
\label{eq:alpha}
    \alpha &= \frac{ 
    2\sqrt{5} \, B_4^4 }{\sqrt{ 
        \left( \sqrt{X^2 + 20\left(B_4^4\right)^2} + X \right)^{2} + 20\left(B_4^4\right)^2}}, \\
      X &= \left(B_{2}^{0} + 20\,B_{4}^{0}\right),\\
    \beta^2 &= 1 -\alpha^2.
\end{align}

\section{LOCALIZED TOTAL ANGULAR-MOMENTUM OPERATORS}
\label{app:Jxyz}
The localized total angular-momentum operators $\hat{J}^{x,y,z}$ at each site $i (j)$ are expressed as
\begin{widetext}
\label{eq:gt71gt72jxjyjz}
\begin{align}\label{eq:jx}
    \hat{J}^{x}= \left(
\begin{array}{cccc}
 0 & -\sqrt{5} \alpha  \beta  & 0 & \frac{1}{2} \sqrt{5} (\alpha -\beta ) (\alpha +\beta ) \\
 -\sqrt{5} \alpha  \beta  & 0 & \frac{1}{2} \sqrt{5} (\alpha -\beta ) (\alpha +\beta ) & 0 \\
 0 & \frac{1}{2} \sqrt{5} (\alpha -\beta ) (\alpha +\beta ) & 0 & \sqrt{5} \alpha  \beta  \\
 \frac{1}{2} \sqrt{5} (\alpha -\beta ) (\alpha +\beta ) & 0 & \sqrt{5} \alpha  \beta  & 0 \\
\end{array}
\right),
\end{align}

\begin{align}\label{eq:jy}
     \hat{J}^{y}=\left(
\begin{array}{cccc}
 0 & i \sqrt{5} \alpha  \beta  & 0 & -\frac{1}{2} i \sqrt{5} (\alpha -\beta ) (\alpha +\beta ) \\
 -i \sqrt{5} \alpha  \beta  & 0 & \frac{1}{2} i \sqrt{5} (\alpha -\beta ) (\alpha +\beta ) & 0 \\
 0 & -\frac{1}{2} i \sqrt{5} (\alpha -\beta ) (\alpha +\beta ) & 0 & -i \sqrt{5} \alpha  \beta  \\
 \frac{1}{2} i \sqrt{5} (\alpha -\beta ) (\alpha +\beta ) & 0 & i \sqrt{5} \alpha  \beta  & 0 \\
\end{array}
\right),
\end{align}

\begin{align}\label{eq:jz}
    \hat{J}^{z}=\left(
\begin{array}{cccc}
 \frac{1}{2} \left(5 \alpha ^2-3 \beta ^2\right) & 0 & 4 \alpha  \beta  & 0 \\
 0 & \frac{1}{2} \left(3 \beta ^2-5 \alpha ^2\right) & 0 & -4 \alpha  \beta  \\
 4 \alpha  \beta  & 0 & \frac{1}{2} \left(5 \beta ^2-3 \alpha ^2\right) & 0 \\
 0 & -4 \alpha  \beta  & 0 & \frac{1}{2} \left(3 \alpha ^2-5 \beta ^2\right) \\
\end{array}
\right),
\end{align}
\end{widetext}
where the basis is given by $\ket{1}$--$\ket{4}$.
On the four crystal-field atomic basis states, the total angular momentum is anisotropic, and the degree of anisotropy can be tuned by varying $\alpha$.

\section{COMPUTATIONAL METHODOLOGY}
\label{app:computational methodology}
Our computational methodology based on the mean-field calculations proceeds as follows.
Simulations are performed on an $N = 6 \times 6$ square lattice with periodic boundary conditions, a system size chosen to be commensurate with the expected ordering wave vectors, $\bm{Q}_1$ and $\bm{Q}_2$. 
The ground state is determined via a self-consistent mean-field approach. 
The iterative procedure begins from an initial magnetic-moment configuration $\{\langle \hat{\bm{J}}_i \rangle\}$. 
For each lattice site $i$, a local effective Hamiltonian is constructed based on mean-field interactions with its neighbors
[\Eq{eq:TotMFhamiltonian}] and then diagonalized to obtain eigenvalues $\left \{ \epsilon_n \right \} $ and eigenvectors $\left \{ \ket{n} \right \} $.
An
updated thermal expectation value for the magnetic moment at site $i$, $\langle \hat{\bm{J}}_i \rangle_{\text{new}}$, is subsequently calculated using Boltzmann statistics over these eigenstates.
The expectation values of the observables $\hat{\mathcal{O}}$ at the temperature $T$ 
are given by
\begin{align}\label{eq:exactDia}
    \langle \hat{\mathcal{O}} \rangle 
    = \frac{1}{Z} \sum_{n} \bra{n} \hat{\mathcal{O}} \ket{n}\exp(- \epsilon_{n}/k_{\rm B}T),
\end{align}
where $Z = \sum_{n} \exp(-\epsilon_{n}/k_{\rm B}T)$ is the partition function,
Here, we set the Boltzmann constant $k_{\rm B} = 1$. 
The temperature $T$ is measured in units of the nearest-neighbor exchange interaction $\mathcal{J}_1$, and all calculations in this study are performed at $T = 0.05$, which is small enough compared to the energy scale of the exchange interactions
and suppresses thermal orbital mixing.

This procedure is repeated for all sites to complete one iteration step~\footnote{The global magnetic-moment configuration is updated with a damping factor to ensure stable convergence.}. 
Iteration continues until both the total free energy and site-resolved magnetic-moment configuration converge to a strict tolerance of $\tau = 10^{-6}$, at which point the system is considered self-consistent.

A key feature of our method is a robust scheme to identify the
ground state and avoid metastable local minima. This is achieved via a massively parallel computational approach, wherein a large and diverse pool of initial magnetic-moment configurations (ansatz) is evolved simultaneously for each parameter point. The initial pool includes:
\begin{itemize}
\item \textit{Standard magnetic orders:} Ferromagnetic, single-$Q$ vertical spiral, double-$Q$ vertical spiral, and double-$Q$ in-plane spiral configurations~\cite{SZLinPhysRevB.93.064430}.
\item \textit{Parameterized state families:} 21 single-$Q$ conical spiral configurations generated by varying the cone angle, and 216 S-SkL configurations generated according to Ref.~\cite{UtesovPhysRevB.103.064414}.
\item \textit{Randomized states:} Approximately 200 distinct random magnetic-moment configurations.
\item \textit{Solutions from external data:} States imported from several separate files of converged solutions to serve as high-quality initial guesses.
\item \textit{Perturbed states:}
Updated initial states generated by adding small random fluctuations, ranging from $-0.5$ to $0.5$, to the imported configurations.
\item \textit{Solution inheritance:} The 300 lowest-energy converged solutions from the preceding magnetic field step, $h_{n-1}$, used as initial states for the current step, $h_n$.
\end{itemize}
The main simulation loop systematically sweeps through the parameter space, sequentially consisting of the external field $h$, 
crystal-field splitting $\Delta$ and superposition coefficient $\alpha$.
For each point, the converged solution with the lowest free energy among all initial trials is identified as the ground state.

\section{PHYSICAL QUANTITIES}
\label{app:physical quantities}
We preliminarily identify the obtained magnetic states by calculating the structure factor. The structure factor is defined as a correlation function of the mean magnetic moment $\langle \hat{\bm{J}}_i \rangle$:
\begin{eqnarray}\label{eq:SSF}
    J^{d}(\bm{q}) &=& \frac{1}{N} \sum_{i,j}^{}  
    \left \langle J^{d}_{i} \right \rangle
     \left \langle  J^{d}_{j} \right \rangle  \exp \left [ i\,  \bm{q}  \cdot   \left (\bm{r}_{i}-\bm{r}_{j}\right)  \right ], \\
    J(\bm{q}) &=&  \sum_{d}J^{d}(\bm{q}),
\end{eqnarray}
where $d$ 
denotes the direction component ($d=x,y,z$) and $N = 36$ is the total number of lattice sites. To further figure out the contribution in each direction, we also decompose the structure factor into the in-plane one $J^x(\bm{q}) + J^y (\bm{q}) \equiv J^{xy}_{\bm{q}}$, and the out-of-plane one, $J^z(\bm{q}) \equiv J^{z}_{\bm{q}}$.

In addition, we calculate the scalar chirality $\chi_i$ around each site and the topological skyrmion number $N_{\text{sk}}$. 
The scalar chirality $\chi_i$ is defined by the scalar triple product of the magnetic moments as $\langle \hat{\bm{J}}_i\rangle \cdot (\langle \hat{\bm{J}}_j\rangle \times \langle \hat{\bm{J}}_k\rangle)$~\cite{XGWenPhysRevB.39.11413}. 
On a square lattice, we modify this definition to evaluate the local scalar chirality at a given site $i$ by summing the scalar triple products over the four triangles surrounding site $i$, and dividing by $2$ to avoid double counting. 
The formula is expressed as
\begin{align}\label{eq:ScalarSpinChirality}
    \chi_{i} = \frac{1}{2} \sum_{\delta,\delta'= \pm1} \delta \delta' \left \langle \hat{\bm{J}}_{i} \right \rangle \cdot \left(  \left \langle \hat{\bm{J}}_{i+\delta\hat{x}} \right \rangle \times  \left \langle \hat{\bm{J}}_{i+\delta'\hat{y}} \right \rangle \right),
\end{align}
where $\hat{x}$ ($\hat{y}$) represents a translation by the lattice constant along the $x$ ($y$) direction. 
The net (nonzero total) scalar chirality, denoted as $\langle \chi \rangle$, is obtained by summing the local scalar chirality over all lattice sites and normalizing by the total number of sites:
\begin{align}\label{eq:netChirality}
\langle \chi \rangle = \frac{1}{N}\sum_i \chi_i.
\end{align}
When a system exhibits net scalar chirality, it often gives rise to novel physical phenomena like the topological Hall effect.

The topological skyrmion number $n_{\text{sk}}$ in the continuum field theory is defined in \Eq{eq:topoNuminIntro}~\footnote{Throughout this paper, the notation $n_{\text{sk}}$ refers to the topological skyrmion number defined in the continuum field theory, whereas $N_{\text{sk}}$ represents its discrete-lattice counterpart.}.
In a discrete-lattice representation, the continuous integral is replaced by a summation over local skyrmion densities (solid angles)~\cite{Rajaraman,braun2012topological}, yielding the total skyrmion number as
\begin{align}\label{eq:skyrmion number}
N_{\text{sk}} = \frac{1}{2}\frac{1}{4\pi} \sum_{i} \Omega_{i}.
\end{align}
The prefactor $\tfrac{1}{2}$ accounts for the presence of two magnetic unit cells within the $6\times6$ unit cell when the SkL or other double-$Q$ 
magnetic textures are realized.
Here, $\Omega_{i} \in [-2\pi,\, 2\pi)$ denotes the local skyrmion density~\cite{berg1981definition}, which is expressed as
\begin{align}\label{eq:S-SkLDensity}
   \Omega_{
   i} =
   \sum_{\delta,\delta'= \pm1} \arctan{\left( \frac{2\delta \delta' \left \langle \hat{\bm{j}}_{
   i} \right \rangle \cdot \left( \left \langle \hat{\bm{j}}_{
   j} \right \rangle \times \left \langle \hat{\bm{j}}_{
   k} \right \rangle \right)}{\left( \left \langle \hat{\bm{j}}_{
   i} \right \rangle + \left \langle \hat{\bm{j}}_{
   j} \right \rangle + \left \langle \hat{\bm{j}}_{
   k} \right \rangle \right)^{2} - 1} \right)}.
\end{align}
where the normalized local moments are defined as
$\left \langle \hat{\bm{j}}_{
i} \right \rangle  = \left \langle \hat{\bm{J}}_{i} \right \rangle/ \abs{\left \langle \hat{\bm{J}}_{i} \right \rangle}$, with neighboring sites $j=i+\delta\hat{x}$ and $k=i+\delta'\hat{y}$.
The quantity $N_{\text{sk}}$ counts the number of times the 
magnetic-moment configuration wraps around the unit sphere along a specific direction, and thus takes integer values.
For instance, $\abs{N_{\text{sk}}} = 1$ corresponds to a single skyrmion contained within one magnetic unit cell.

Furthermore, we evaluate the net magnetization, defined as the vector sum of all mean magnetic moments divided by the total number of sites $N$:
\begin{align}
\langle \hat{\bm{J}} \rangle = \frac{1}{N} \sum_{i} \langle \hat{\bm{J}}_i \rangle.
\end{align}
To quantitatively characterize its magnitude, we consider the norm of $\langle \hat{\bm{J}} \rangle$, denoted as $\abs{\langle \hat{\bm{J}} \rangle}$.

\section{PHASE CLASSIFICATION CRITERIA}\label{app:phaseclassification}
We follow the definitions of the 1$Q$~CS state, the S-SkL state, and other 2$Q$~I--VII states given in Ref.~\cite{ZhaHayamiPhysRevB.111.165155} and analyze the decomposed structure factors $J_{\bm{q}}^{xy}$ and $J_{\bm{q}}^{z}$ with a numerical precision of $\tau = 10^{-6}$. 
The decomposed structure factors are numerically classified according to the following criteria: 
Values in the range of $(10^{-2},\,\tau)$ are regarded as $\ll 1$ but not exactly zero, whereas values within $(\tau,\,0)$ are treated as zero. 
If the absolute difference between $J^{\zeta}_{\bm{Q}_1}$ and $J^{\zeta}_{\bm{Q}_2}$ (or between $J^{\zeta}_{\bm{Q}'_1}$ and $J^{\zeta}_{\bm{Q}'_2}$) falls within $(10^{-2},\,\tau)$, we consider them approximately equal, i.e., 
$J^{\zeta}_{\bm{Q}_1} \simeq J^{\zeta}_{\bm{Q}_2}$ (or $J^{\zeta}_{\bm{Q}'_1} \simeq J^{\zeta}_{\bm{Q}'_2}$). 
When the difference lies within $(\tau,\,0)$, they are treated as identical, 
$J^{\zeta}_{\bm{Q}_1} = J^{\zeta}_{\bm{Q}_2}$ (or $J^{\zeta}_{\bm{Q}'_1} = J^{\zeta}_{\bm{Q}'_2}$). 

Due to the classification accuracy used here 
differing from that in Ref.~\cite{ZhaHayamiPhysRevB.111.165155}, the definitions of the multi-$Q$ states may not be exactly the same. 
In certain cases, some phases evolve continuously as the external magnetic field $h$ or the crystal-field splitting $\Delta$ varies; 
For example, a finite component of the decomposed structure factor may gradually approach zero, or two close components may merge into one.
Since such changes represent continuous deformations, we classify them manually as belonging to the same phase, rather than strictly applying the numerical criteria described above.

\section{SNAPSHOTS OF THE EIGHT REPRESENTATIVE STATES IN THE PREVIOUS STUDY}
\label{app:states}
\begin{figure*}[t!]
    \begin{center}
    \includegraphics[width=0.85 \hsize]{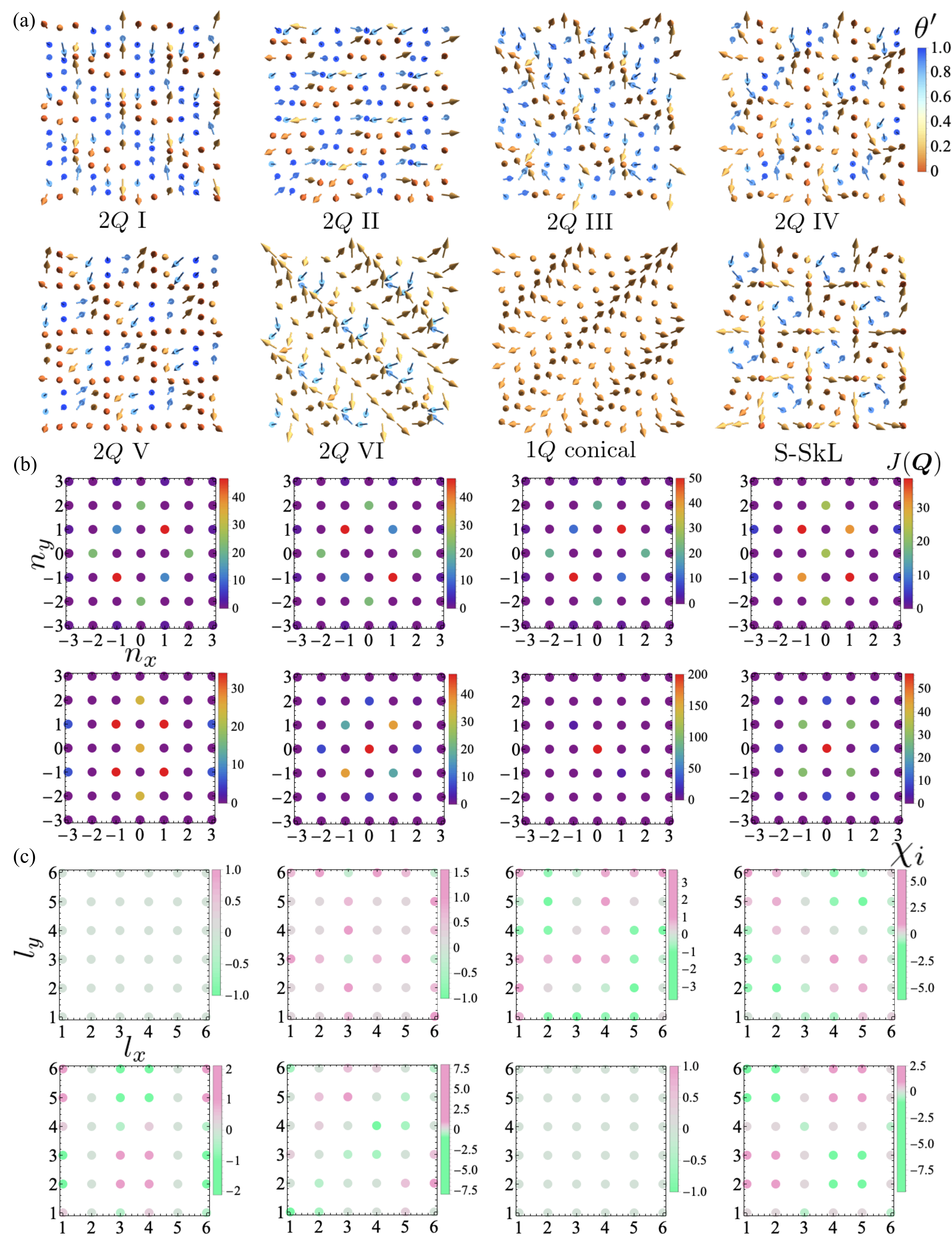}
    \caption{Real-space and momentum-space characteristics of the magnetic phases 2$Q$~I--VI, 1$Q$~CS and S-SkL stabilized either only in $\Delta>0$ or in both $\Delta>0$ and $\Delta<0$ regions, depending on parameters~\cite{ZhaHayamiPhysRevB.111.165155}.
    Panels (a)--(c) show (a) the three-dimensional magnetic-moment configurations in a $6 \times 6$ unit cell, where the magnetic moments are drawn at each site and normalized in length for clarity; 
    the color scale represents the normalized polar angle $\theta' = \theta / \pi$, varying continuously from 0 (north pole) to 1 (south pole); 
    (b) the structure-factor distributions $J(\bm{q})$ in momentum space, where $n_x$ and $n_y$ denote multiples of $2\pi / 6$ with $-3 \le n_x, n_y \le 3$ in the first Brillouin zone; 
    and (c) the local scalar chirality $\chi_i$ within the $6 \times 6$ unit cell.
    }
    \label{fig_RealspaceSfChirality_previous}
    \end{center}
\end{figure*}

To avoid redundancy with Ref.~\cite{ZhaHayamiPhysRevB.111.165155} while maintaining a self-contained presentation, we present in Fig.~\ref{fig_RealspaceSfChirality_previous} the snapshots of the eight representative states that were already documented there (the 1$Q$ conical state, the S-SkL, and the 2$Q$~I--VI states), including their real-space magnetic-moment configurations, structure-factor maps, and local scalar-chirality maps. 
We instead focus the main part of this paper on emergent phases 
in the present work, particularly those unique to $\Delta<0$ and/or to the extended parameter ranges.

\section{MULTIPOLAR (TENSORIAL) EXCHANGE BEYOND THE BILINEAR FORM}
\label{app:tensor_exchange}
As discussed in Sec.~\ref{sec:model}, the strong relativistic spin--orbit coupling together with the tetragonal crystal-field effects projects the $\ket{5/2, J^z}$ multiplet onto two low-lying Kramers doublets. 
The superposition coefficient $\alpha$ controls the relative weights of $\ket{\tfrac{5}{2},\pm \tfrac{5}{2}}$ and $\ket{\tfrac{5}{2},\mp \tfrac{3}{2}}$, and thereby determines the anisotropic matrix elements of the projected $\hat{\bm{J}}$ operators. 
Strictly speaking, for ions with unquenched orbital angular momentum ($\bm{L}\neq \bm{0}$), the intersite exchange interaction
is not a simple scalar product but takes a multipolar and anisotropic tensor form,
\begin{align} \label{eq:iwahara} 
\sum_{kq,k'q'} \mathcal{J}_{kq,k'q'}
\frac{\hat{O}_{k}^{q}(\hat{\bm{J}}_1) \hat{O}_{k'}^{q'}(\hat{\bm{J}}_2)}{O_k^0(J_1) O_{k'}^0(J_2)},
\end{align}
as rigorously demonstrated by Iwahara and Chibotaru~\cite{IwaharaPhysRevB.91.174438}.
Here, $J_1$ and $J_2$ denote the total-angular-momentum quantum numbers on sites 1 and 2 (not exchange constants), and $O_k^0(J_i)$ are scalars obtained by replacing $\hat{\bm{J}}_i^{\,2}$ and $\hat{J}^{z}_{i}$ in $O_k^0(\hat{\bm{J}}_i)$ with the eigenvalues $J_i(J_i+1)$ and $J_i$, respectively~\cite{IwaharaPhysRevB.91.174438}.
In the present study, however, we adopt the simplified bilinear form of the exchange Hamiltonian,
$-\mathcal{J} \bm{J}_1 \cdot \bm{J}_2$.
Since the anisotropy has already been incorporated at the single-ion level through the crystal-field orbital wave functions, the effective exchange acquires a strong spin dependence without the need to explicitly introduce tensorial exchange parameters. 
This approximation enables us to focus on the key roles of multiorbital effects and higher-harmonic wave vectors, while a full tensorial exchange treatment~\cite{IwaharaPhysRevB.91.174438} is deferred to future work.

\bibliography{apssamp}

\end{document}